\documentclass[pdflatex,sn-mathphys-num]{sn-jnl}

\usepackage{graphicx}
\usepackage{amsmath,amssymb,amsfonts}
\usepackage{xcolor}
\usepackage{booktabs}
\usepackage{subcaption}
\usepackage{tikz}
\usepackage{mdframed}
\usepackage{braket}
\usepackage{epigraph}
\usepackage{algorithm}
\usepackage{algpseudocode}
\usetikzlibrary{arrows.meta,positioning,decorations.pathreplacing}

\usepackage{placeins}

\usepackage[normalem]{ulem}

\usepackage{etoolbox}
\makeatletter
\patchcmd{\@maketitle}{Contributing authors:\ \authemail}{E-mail:\ \authemail}{}{}
\renewcommand{\email}[1]{%
  \global\advance\emailcnt by 1\relax
  \ifnum\emailcnt=1\relax
    \g@addto@macro\authemail{\textcolor{blue}{#1}}%
  \else
    \g@addto@macro\authemail{, \textcolor{blue}{#1}}%
  \fi}
\makeatother

\begin{document}

\title{Quantum simulation of real-time bosonic dynamics \break with binary and unary position encodings \break on superconducting and trapped-ion processors}

\author[1]{\fnm{Masanori} \sur{Hanada}}\email{m.hanada@qmul.ac.uk}
\author[2]{\fnm{Jiwoo} \sur{Seo}}\email{tjwldn1234@khu.ac.kr}
\author[2,3,4,5]{\fnm{Seok-Kyun} \sur{Son}}\email{skson@khu.ac.kr}
\author[2,3,4,5]{\fnm{Junggi} \sur{Yoon}}\email{junggi.yoon@khu.ac.kr}

\affil[1]{\orgdiv{School of Mathematical Sciences}, \orgname{Queen Mary University of London}, \orgaddress{\street{Mile End Road}, \city{London}, \postcode{E1 4NS}, \country{United Kingdom}}}
\affil[2]{\orgdiv{Department of Physics, College of Science}, \orgname{Kyung Hee University}, \orgaddress{\street{26 Kyunghee-daero, Dongdaemun-gu}, \city{Seoul}, \postcode{02447}, \country{Republic of Korea}}}
\affil[3]{\orgdiv{International Research Institute for Quantum Materials}, \orgname{Kyung Hee University}, \orgaddress{\street{26 Kyunghee-daero, Dongdaemun-gu}, \city{Seoul}, \postcode{02447}, \country{Republic of Korea}}}
\affil[4]{\orgdiv{Neutral Atom-based Quantum Application Lab}, \orgname{Kyung Hee University}, \orgaddress{\street{26 Kyunghee-daero, Dongdaemun-gu}, \city{Seoul}, \postcode{02447}, \country{Republic of Korea}}}
\affil[5]{\orgdiv{Research Institute for Basic Sciences}, \orgname{Kyung Hee University}, \orgaddress{\street{26 Kyunghee-daero, Dongdaemun-gu}, \city{Seoul}, \postcode{02447}, \country{Republic of Korea}}}

\abstract{ Simulating a bosonic degree of freedom on qubit hardware requires
discretizing its position space, and the qubit encoding of the resulting
lattice determines the circuits that realize its real-time dynamics. We
compare binary and unary encodings of a position lattice on
superconducting (IBM) and trapped-ion (IonQ) processors. Binary encoding
applies the kinetic term through the quantum Fourier transform; unary
encoding uses one qubit per site, single-qubit rotations for the potential
and excitation-conserving exchange gates for the kinetic term. For a single
degree of freedom the unary dynamics is a free-fermion problem, and its
product formula compresses exactly into a nearest-neighbor circuit of fixed
size. With about two hundred two-qubit gates, these circuits resolve
tunneling in a symmetric double well on both processors, with the
right-well probability falling from 0.97 to 0.06 on IBM and from 0.96 to
0.12 on IonQ, whereas binary product-formula circuits with several thousand
two-qubit gates do not reproduce the transfer. Noiseless emulation
indicates that this contrast reflects the compilation as well as the
encoding: without the fixed-size optimization, a unary product formula that
captures the transfer also requires several thousand two-qubit gates.
Post-selection on the single-excitation subspace removes leakage, and
because every energy eigenstate of the symmetric grid populates both wells
equally, the measured population reversal bounds the coherence between
energy eigenstates from below. Coupled-oscillator benchmarks, where
inter-register interactions prevent this exact compression, show that the
preferred encoding depends on the grid size and the evolution time.}


\maketitle
\newpage

\section{Introduction}\label{sec:intro}

{
\setlength{\epigraphwidth}{0.7\textwidth}
\epigraph{\itshape
``The career of a young theoretical physicist consists of treating
the harmonic oscillator in ever-increasing levels of abstraction.''
}{--- \textsc{Sidney Coleman}}
}

The harmonic oscillator is the simplest quantum system and, at the same
time, the one from which much of modern physics has been built.
Planck's resolution of the black-body problem was a statement about
oscillators~\cite{Planck:1901oar}; Bose and Einstein, extending the same
counting to material particles, found that a gas of oscillator modes
condenses~\cite{Bose:1924mk,Einstein:1924}, and Feynman later showed that
superfluidity is this condensation at work in an interacting
liquid~\cite{Feynman:1953zza}.
Born, Heisenberg and Jordan, Dirac, and Heisenberg and Pauli quantized
the electromagnetic field by treating each mode as an oscillator, and
quantum field theory was born~\cite{Born:1926uzf,Dirac:1927,Heisenberg:1929xj}.
More recently, a lattice of coupled oscillators was enough for Srednicki
to uncover the area law of entanglement entropy~\cite{Srednicki:1993im},
and the free-field limit of Yang--Mills theory on a sphere---again an
oscillator gas---exhibits the Hagedorn growth that signals
deconfinement~\cite{Sundborg:1999ue,Aharony:2003sx}.
In the orbifold lattice formulation, Hamiltonian Yang--Mills theory is
a system of interacting oscillators with the canonical structure
$[\hat x_a,\hat p_b]=\mathrm{i}\delta_{ab}$~\cite{Bergner:2024qjl}.

Oscillator-based position registers therefore provide building blocks
for Hamiltonian lattice Yang--Mills simulations~\cite{Halimeh:2024bth};
extending the present circuits to those theories additionally requires
their interactions and constraints. It is therefore natural to ask
whether the harmonic oscillator and its simplest anharmonic
deformations can be simulated faithfully on the quantum devices
available today. By a bosonic degree of freedom we mean a canonical
pair $[\hat x,\hat p]=\mathrm{i}$ with continuous spectrum, such as a
vibrational mode, the amplitude of a lattice field at one site, or a
matrix element in matrix quantum mechanics. We consider $K$ such
degrees of freedom with Hamiltonian
\begin{equation}
  \hat H=\frac12\sum_{a=1}^{K}\hat p_a^{2}+V(\hat x_1,\ldots,\hat x_K),
  \qquad [\hat x_a,\hat p_b]=\mathrm{i}\delta_{ab},\quad a,b=1,\ldots,K.
  \label{eq:HK}
\end{equation}
We focus on a single degree of freedom, $K=1$, and then examine
benchmarks with an interaction between two degrees of freedom.

Digital simulation of continuous bosonic degrees of freedom introduces errors associated with the finite coordinate range, spatial discretization, product-formula approximation, and hardware implementation. The harmonic oscillator provides a useful benchmark because its analytic solution allows these effects to be assessed against controlled references. This becomes particularly important for anharmonic and interacting systems, where obtaining independent continuum solutions is generally more demanding.

For the harmonic oscillator, continuum, exact-grid, and ideal product-formula references provide a hierarchy for assessing simulation accuracy. Their differences distinguish spatial discretization and product-formula errors, while deviations between QPU measurements and ideal-circuit predictions characterize the overall implementation error. Separating finite-domain effects from grid-spacing errors requires a systematic study in which the domain size and lattice spacing are varied independently.
Moreover, because interacting bosonic theories, including Hamiltonian lattice Yang–Mills theory, are built from coupled anharmonic oscillators, an encoding that fails to reproduce a single oscillator faithfully cannot be expected to succeed for them.

An important foundation for position-space digitization was established by Jordan, Lee and Preskill~\cite{Jordan:2012xnu}, who represented the continuous field amplitude at each spatial lattice site by a finite set of values stored in a qubit register. Their quantum algorithm for scalar-field scattering provided a framework for controlling the errors introduced by truncating and discretizing the field. Macridin et al.~\cite{Macridin:2018gdw,Macridin:2018oli} developed a position-grid representation of bosonic modes in electron–phonon systems, showing how low-energy oscillator states can be accurately represented on a finite grid. Their evolution circuits implement the potential through position-dependent phases and the kinetic term through a quantum Fourier transform, momentum-dependent phases, and an inverse transform. Klco and Savage~\cite{Klco:2018zqz} systematically examined scalar-field digitization, relating the field range and grid spacing to digitization errors and analyzing the associated qubit and gate requirements.

Complementary studies have explored bosonic simulations on noisy quantum hardware. Yeter-Aydeniz et al.~\cite{Yeter-Aydeniz:2018mix} calculated low-lying energy levels of a scalar field theory using a variational algorithm, while Klco and Savage~\cite{Klco:2019xro} demonstrated the preparation of localized wavefunctions in a discretized field basis on IBM devices. These studies established hardware implementations of spectral calculations and state preparation. Real-time evolution poses an additional challenge: the prepared wave packet must undergo kinetic and potential operations while preserving the features of its dynamics as circuit errors accumulate.

Unary encoding offers an alternative representation of the discretized
coordinate: an $N$-site grid is stored in $N$ qubits carrying a single
excitation, whose location labels the basis state. This construction,
often called the direct or one-hot mapping, has been used to represent
bosonic Fock or vibrational levels~\cite{Sawaya:2020bky,Ollitrault:2020kms}, including
in studies of vibrational structure and spin--boson
dynamics~\cite{Ollitrault:2020kms,Miessen:2021jdn}. Here each qubit
instead labels a site of the position grid, and the excitation specifies
the position of the encoded degree of freedom. The potential is then
implemented by local $Z$ rotations, and a finite-difference kinetic
operator becomes excitation-conserving hopping between neighboring
sites. Unary encoding therefore uses more qubits than binary encoding, but its
gates are local. For a single degree of freedom, the unary Hamiltonian is
quadratic in Jordan--Wigner fermions. Local gate identities
can therefore reduce repeated Trotter steps to a circuit
whose size is independent of their number, while preserving
the product-formula unitary on the single-excitation
subspace~\cite{Camps_2022}. The resulting triangular
circuit is the Givens-rotation decomposition familiar from linear
optics~\cite{Reck:1994ujl,Clements:2016rkw} and from free-fermion state preparation~\cite{Kivlichan:2018bqq,Jiang:2017pyp}. Since the dynamics of a
single degree of freedom can also be computed classically, our aim is
not a computational advantage but a controlled comparison of encodings
and compilation strategies under hardware noise.

We implement binary and unary position-space dynamics on IBM Quantum's 
\texttt{ibm\_phoenix} (Nighthawk r2) and IonQ Forte (\texttt{forte-1}). The harmonic
oscillator benchmarks wave-packet motion and spreading, and the
symmetric double well probes population transfer governed by the
splitting of its lowest energy doublet. We further consider bilinear and
quartic couplings between two degrees of freedom to examine the
implementations in the presence of inter-register interactions. We compare the measured dynamics with three references: the continuum
evolution, the exact evolution under the discretized Hamiltonian, and
the noiseless product-formula circuit. The optimized unary circuit preserves the product-formula unitary,
and its size bound is independent of the number of steps. Thus, at
fixed evolution time, the time step can be reduced without increasing
that bound. At any chosen finite step size, residual product-formula
error must be distinguished from spatial discretization and hardware
effects. For the fourteen-site double well, the optimized unary circuits use 208
two-qubit gates on IBM Phoenix and 182 on IonQ Forte up to $t=30.4$,
whereas the binary product-formula circuits require $7\,186$ and $3\,320$
up to $t=26.4$. The unary data on both processors resolve the transfer
of the wave packet between the wells, with the right-well probability
falling from $0.97$ to $0.06$ (IBM) and from $0.96$ to $0.12$ (IonQ),
while the binary data do not reproduce the transfer: on IBM the
measured position distribution is nearly uniform over the grid at late
times, while on IonQ the right-well preference persists. Circuit
resources also depend on the chosen compilation. The noiseless
product-formula baselines in
Appendix~\ref{app:compilation} show that resolving the transfer
can require several thousand two-qubit gates in either
implementation. Excitation-number post-selection rejects measured outcomes outside
the unary subspace. Because every energy eigenstate of the symmetric fourteen-site grid populates the two wells equally, the measured population reversal bounds the coherence between energy eigenstates from below, with 95\% lower confidence limits of 0.85 (IBM) and 0.62 (IonQ) after the first transfer. The coupled-oscillator benchmarks show that on small grids the binary encoding remains competitive, so the preferred encoding depends on the grid size, the evolution time and the connectivity of the device.

\subsubsection*{Relation to other work}
The same lattice structures appear in digital simulations of compact
lattice gauge theories. A periodic lattice of $N$ sites has a cyclic
$\mathbb{Z}_N$ translation symmetry, and its sites may be labeled by
$\theta_j=2\pi j/N$, with $j=0,\ldots,N-1$. Such periodic grids are also
used to represent compact angular coordinates in Abelian lattice gauge
theories. In a coordinate representation, potential terms are diagonal
and can be implemented through coordinate-dependent phase operations.
The kinetic evolution involves structures closely related to those used
here. In our binary implementation, the wavefunction is transformed to
the Fourier basis, phases determined by the momentum-squared spectrum
are applied, and the state is transformed back to the position basis.
This is analogous to implementing the electric-energy term in the
Fourier-transformed electric basis of an Abelian gauge theory. In our
unary implementation, the second spatial derivative is approximated
directly on the position grid by finite differences. The resulting
coupling between neighboring grid sites is implemented using local,
excitation-preserving two-qubit rotations.

Discrete $\mathbb{Z}_N$ approximations to $U(1)$ have been
studied~\cite{Notarnicola2015}, and general methods for digital
gauge-theory simulation have been developed~\cite{LammLawrenceYamauchi2019}.
Magnetic-basis representations designed to remain efficient at weak
coupling have also been proposed~\cite{Haase2021,BauerGrabowska2023}.
Single-plaquette quantum-simulation protocols and their resource
requirements were studied by Paulson et al.~\cite{Paulson2021}, while
subsequent trapped-ion qudit experiments demonstrated gauge-field
truncation effects and real-time gauge-theory dynamics~\cite{Meth2025}.

The present setting involves a noncompact coordinate $x\in\mathbb{R}$.
Unlike an angular coordinate with a fixed physical periodicity, its
finite-grid representation requires a truncation length $L$ and a grid
spacing $\Delta x$, whose effects must be controlled separately.
Noncompact coordinate variables also arise in the orbifold lattice
formulation of Yang--Mills theory and in matrix quantum
mechanics~\cite{Bergner:2024qjl,Halimeh:2024bth}, providing a motivation
for studying oscillator-based building blocks. We focus on the real-time
dynamics of localized wave packets, for which the grid must be wide
enough to contain the evolving spatial distribution and fine enough to
resolve its structure. The harmonic oscillator and double well provide
elementary models for examining how a finite position-grid
representation captures these dynamics.


\section{Bosonic Models and Lattice Encoding}\label{sec:lattice}

\subsection{Models and Initial States}

We consider a single bosonic degree of freedom governed by the
Hamiltonian
\begin{equation}
    \hat{H}=\frac{\hat{p}^{\,2}}{2}+V(\hat{x}),
\end{equation}
in units where $\hbar=m=1$.
We investigate two potentials: the harmonic oscillator (HO)
and the symmetric double-well (DW) potential,
\begin{equation}
    V_{\mathrm{HO}}(x)=\frac{k}{2}x^2,
    \qquad
    V_{\mathrm{DW}}(x)
    =\frac{V_0}{a^4}(x^2-a^2)^2.
    \label{eq:potentials}
\end{equation}
The harmonic oscillator has frequency $\omega=\sqrt{k}$,
whereas the double-well potential has two minima at
$x=\pm a$, separated by a barrier of height $V_0$.

\paragraph{Harmonic Oscillator}

For the harmonic oscillator, we consider an initial
Gaussian wave packet centered at $x_0$ with zero mean
momentum,
\begin{equation}
    \psi_{\mathrm{HO}}(x,0)
    =\frac{1}{(\pi s^2)^{1/4}}
    \exp\left[-\frac{(x-x_0)^2}{2s^2}\right],
    \label{eq:ho_initial}
\end{equation}
where $s$ characterizes the wave-packet width.
The exact continuum dynamics are given by
\begin{align}
    \langle x\rangle(t)
    &=x_0\cos(\omega t),\label{eq:ho-moments}\\
    \mathrm{Var}(x)(t)
    &=\frac{s^2}{2}\cos^2(\omega t)
    +\frac{1}{2\omega^2s^2}\sin^2(\omega t).
\end{align}
For $s^2=1/\omega$, the Gaussian corresponds to
a coherent state, whose variance remains constant,
$\mathrm{Var}(x)=1/(2\omega)$. The energy of the initial state is
\begin{equation*}
  E=\frac{1}{4s^2}+\frac{k\,s^2}{4}+\frac{k\,x_0^2}{2}
  =\frac{1}{4s^2}+\frac{\omega^2 s^2}{4}+\frac{\omega^2x_0^2}{2}.
\end{equation*}

\paragraph{Double-Well Potential}

We consider the symmetric double-well potential
\begin{equation}
    V_{\mathrm{DW}}(x)
    =\frac{V_0}{a^4}(x^2-a^2)^2,
    \qquad V_0,\;a>0,
    \label{eq:dw_potential}
\end{equation}
with minima at $x=\pm a$ and barrier height $V_0$.
Let $\phi_0(x)$ and $\phi_1(x)$ be the normalized
ground and first excited eigenstates, with energies
$E_0$ and $E_1$, respectively. Reflection symmetry
allows these eigenfunctions to be chosen real,
with $\phi_0$ even and $\phi_1$ odd.

For sufficiently deep wells, the lowest two states
form a nearly degenerate doublet. Choosing the sign
of $\phi_1$ so that the eigenfunctions interfere
constructively in the right well, we prepare
\begin{equation}
    \ket{\psi_R}
    =\frac{\ket{\phi_0}+\ket{\phi_1}}{\sqrt{2}},
    \label{eq:dw_initial}
\end{equation}
which is localized predominantly in the right well.
In units with $\hbar=1$, its evolution is
\begin{equation}
    \ket{\psi_R(t)}
    =\frac{e^{-\mathrm{i} E_0t}}{\sqrt{2}}
    \left(
        \ket{\phi_0}
        +e^{-\mathrm{i}\Delta E t}\ket{\phi_1}
    \right),
    \qquad \Delta E=E_1-E_0.
    \label{eq:dw_evolution}
\end{equation}
The relative phase between the two eigenstates
drives oscillations between the wells, with a
characteristic transfer time
$t_{\mathrm{tun}}=\pi/\Delta E$ and a return
period $T_{\mathrm{osc}}=2\pi/\Delta E$.

\begin{figure}[htbp]
    \centering
\includegraphics[
    width=1.0\linewidth,
    height=0.40\textheight,
    keepaspectratio
]{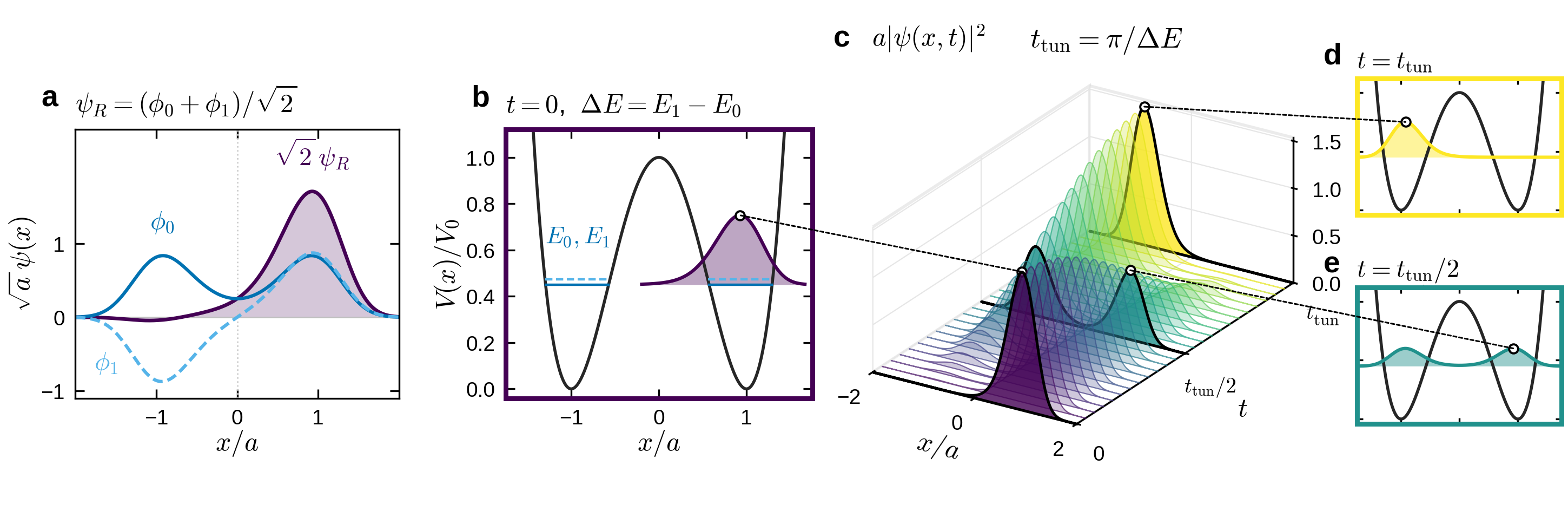}
    \caption{\textbf{Tunneling in a symmetric double-well potential.} (a) The lowest even and odd eigenstates combine into the right-localized state $\psi_R=(\phi_0+\phi_1)/\sqrt2$. (b) The potential with the doublet energies and the initial distribution. (c) Evolution of $a|\psi(x,t)|^2$ up to $t_{\mathrm{tun}}=\pi/\Delta E$. (d,e) Distributions at $t=t_{\mathrm{tun}}$ and $t_{\mathrm{tun}}/2$. Distributions overlaid on the potential are rescaled and offset for clarity.}
    \label{fig:dw_overview}
\end{figure}

Figure~\ref{fig:dw_overview} illustrates the construction of the
right-localized initial state and the resulting probability transfer
between the wells.

\subsection{Truncation and discretization}\label{sec:trunc}

We restrict the spatial coordinate to a finite interval
$x\in[-L,L]$ and discretize it into $N$ lattice sites.
Throughout this work, the physical lattice sites are labeled
by $j=1,2,\ldots,N$, independently of the encoding scheme.
The position of the $j$-th site is defined as
\begin{equation}
    x_j=\left(j-\frac{N+1}{2}\right)\Delta x,
    \qquad
    \sum_{j=1}^{N}x_j=0,
    \label{eq:position}
\end{equation}
where the lattice spacing $\Delta x$ depends on the
boundary condition.

For periodic boundary conditions (PBC), we choose
\begin{equation}
    \Delta x=\frac{2L}{N},
    \qquad
    \psi_{j+N}=\psi_j,
    \label{eq:grid_pbc}
\end{equation}
where sites separated by $N$ lattice spacings are
identified periodically.

For Dirichlet boundary conditions (DBC), we instead choose
\begin{equation}
    \Delta x=\frac{2L}{N+1},
    \qquad
    x_0=-L,\quad x_{N+1}=L,
    \qquad
    \psi_0=\psi_{N+1}=0.
    \label{eq:grid_dbc}
\end{equation}
In this case, the $N$ lattice sites lie strictly
inside the interval, with the wavefunction vanishing
at both boundaries.

The position basis $\{\ket{x_j}\}_{j=1}^{N}$ is
orthonormal, and the position operator is defined by
\begin{equation}
    \hat{x}=\sum_{j=1}^{N}x_j\ket{x_j}\bra{x_j},
    \qquad
    \hat{x}\ket{x_j}=x_j\ket{x_j}.
    \label{eq:position_operator}
\end{equation}
A normalized quantum state can then be expressed as
\begin{equation}
    \ket{\psi}=\sum_{j=1}^{N}\psi_j\ket{x_j},
    \qquad
    \sum_{j=1}^{N}|\psi_j|^2=1,
    \qquad
    \braket{x_i|x_j}=\delta_{ij}.
    \label{eq:discrete_wavefunction}
\end{equation}

The site probability is $P_j= \left|\psi_j\right|^2$, with $\sum_jP_j=1$. It is a probability per retained site, rather than
the continuum density  $\left|\psi(x)\right|^2$. When a continuum profile is sampled to construct a register state, the
sampled amplitudes are normalized with the discrete sum in Eq.~\eqref{eq:discrete_wavefunction}.

Although the physical site index starts at $j=1$,
its representation depends on the encoding.
In binary encoding, the site $\ket{x_j}$ is mapped
to the computational basis state $\ket{j-1}$,
whereas in unary encoding it corresponds to a
single excitation at the $j$-th site.

At matching $N$ and coordinates, either encoding can represent the same $N$-dimensional position space. The implementations studied here nevertheless use different boundary conditions and kinetic matrices, as specified below, and comparisons between them must take these differences into account.

\begin{figure}[htbp]
\centering
\begin{tikzpicture}[x=1cm,y=1cm,>=Latex,line cap=round,line join=round,
    dot/.style={circle,fill=black,inner sep=1.3pt},
    odot/.style={circle,draw=black,fill=white,inner sep=1.3pt},
    lab/.style={font=\small,align=center},
    tit/.style={font=\small\bfseries,align=center}]
  \begin{scope}
    \node[font=\small,anchor=base] at (0,1.85) {(a)\quad\textbf{Periodic boundary}};
    \draw[thick] (0,0) circle (1.4);
    \foreach \a in {0,36,...,324}{\node[dot] at (\a:1.4) {};}
    \node[lab,below=6pt] at (0,-1.4) {$N$ sites, $\Delta x=2L/N$};
  \end{scope}
  \begin{scope}[shift={(3.4,0)}]
    \node[font=\small,anchor=base] at (3,1.85) {(b)\quad\textbf{Dirichlet boundary}};
    \draw[thick] (0,0) -- (6,0);
    \draw[line width=1.2pt] (0,-0.4) -- (0,0.4);
    \draw[line width=1.2pt] (6,-0.4) -- (6,0.4);
    \foreach \x in {0.6,1.2,...,5.5}{\node[dot] at (\x,0) {};}
    \node[odot] at (0,0) {};
    \node[odot] at (6,0) {};
    \draw[<->,thick] (1.2,0.4) -- node[above,font=\footnotesize]{$\Delta x=2L/(N+1)$} (1.8,0.4);
    \draw[decorate,decoration={brace,mirror,amplitude=6pt}] (0.6,-0.3) -- node[below=8pt,lab]{$N$ sites} (5.4,-0.3);
    \node[lab,below=2pt] at (0,-0.4) {$-L$};
    \node[lab,below=2pt] at (6,-0.4) {$L$};
  \end{scope}
\end{tikzpicture}
\caption{The interval $[-L,L]$ is discretized with $N$ sites.
(a)~Periodic boundary condition: the endpoints are identified, giving a closed lattice with spacing $\Delta x=2L/N$.
(b)~Dirichlet boundary condition: $N$ interior sites with spacing $\Delta x=2L/(N+1)$; the endpoints $\pm L$ (open circles) are fixed walls.}
\label{fig:lattice}
\end{figure}
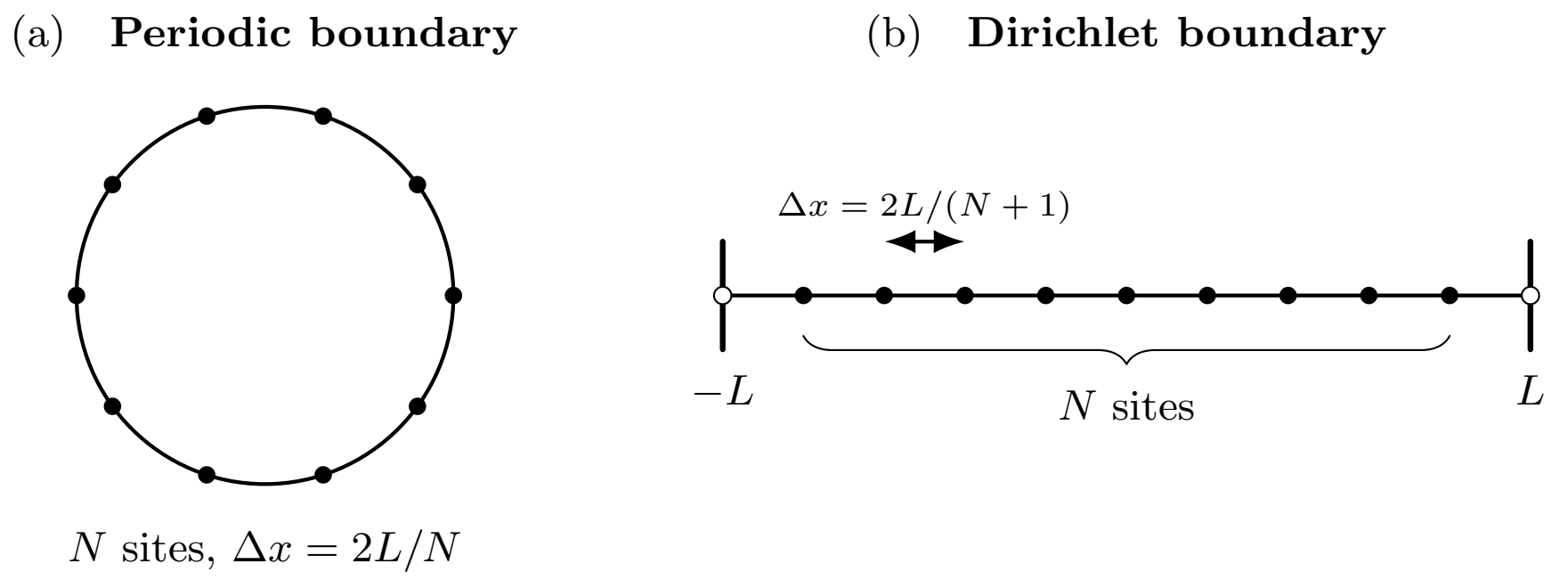

Figure~\ref{fig:lattice} shows the lattices with boundary conditions.

\subsection{Binary and Unary Encoding}
 The grid of Eq.~\eqref{eq:position} can be encoded in two ways, \emph{binary} and \emph{unary}. For a given number of qubits $Q$, the binary encoding uses $N=2^Q$ sites, and the unary encoding uses $N=Q$ sites. The binary encoding is more qubit-efficient, with the potential
and kinetic operators diagonal in the position and momentum
bases, respectively, whereas the unary encoding provides
a simpler local structure for both operators.

\subsubsection{Position Operator Encoding}

\paragraph{Binary}
In the binary encoding, the $N=2^Q$ position-grid points
are represented using $Q$ qubits. We adopt periodic boundary
conditions, with the position coordinates defined as in
Sec.~\ref{sec:trunc}. Each position-grid state is mapped
to a computational basis state labeled by the corresponding
zero-based grid index $j-1$,
\begin{align}
\label{eq:bin_pos_basis}
\ket{x_j}
&\longmapsto \ket{j-1}
=\ket{b_{Q-1}b_{Q-2}\cdots b_0},\\
j-1
&=\sum_{\ell=0}^{Q-1}2^\ell b_\ell,
\qquad b_\ell\in\{0,1\},
\end{align}
where $j=1,2,\ldots,N$. The shift by one accounts for the
different indexing conventions: the position-grid index
starts at $j=1$, whereas the computational basis is labeled
by integers from $0$ to $N-1$.

The decimal-index operator can be expressed in terms of
single-qubit Pauli-$Z$ operators as
\begin{align}
\label{eq:decimal}
\hat N_{\mathrm{dec}}
&=\sum_{\ell=0}^{Q-1}
2^\ell\frac{I-Z_\ell}{2},\\
Z_\ell
&=I^{\otimes(Q-\ell-1)}
\otimes Z\otimes I^{\otimes\ell},
\end{align}
which satisfies
$\hat N_{\mathrm{dec}}\ket{j-1}
=(j-1)\ket{j-1}$.
Consequently, the position operator becomes
\begin{align}
\label{eq:binary_pos}
\hat{x}
&=\Delta x\left(
\hat N_{\mathrm{dec}}-\frac{N-1}{2}I
\right)\notag\\
&=-\frac{\Delta x}{2}
\sum_{\ell=0}^{Q-1}2^\ell Z_\ell,
\end{align}
where we have used
$\sum_{\ell=0}^{Q-1}2^\ell=2^Q-1=N-1$.

Since the position operator is a weighted sum of mutually
commuting Pauli-$Z$ operators, any polynomial potential
$V(\hat{x})$ can be expressed as a linear combination of
Pauli-$Z$ strings and is therefore diagonal in the
computational basis. In particular, the $n$-th power of
the position operator is given by
\begin{align}
\label{eq:binary_x_power}
\hat{x}^n
=
\left(-\frac{\Delta x}{2}\right)^n
\sum_{i_1,\ldots,i_n=0}^{Q-1}
2^{i_1+\cdots+i_n}
Z_{i_1}\cdots Z_{i_n}.
\end{align}

Since the Pauli-$Z$ operators mutually commute and satisfy
$Z_i^2=I$, repeated indices can be eliminated in pairs.
Consequently, $\hat{x}^n$ reduces to a linear combination
of Pauli-$Z$ strings with weights at most $\min(n,Q)$
and with the same parity as $n$.
Even when the polynomial degree exceeds the number of
qubits ($n>Q$), no Pauli string contains more than $Q$
non-identity operators. Further explicit expansions
are provided in Appendix~\ref{bianry_pos}.

For example, the quadratic position operator, which
determines the harmonic oscillator potential, is
\begin{align}
\label{eq:binary_x_squared}
\hat{x}^2
={}&\frac{\Delta x^2(4^Q-1)}{12}I
\notag\\
&+\frac{\Delta x^2}{2}
\sum_{0\leq i<j<Q}
2^{i+j}Z_iZ_j.
\end{align}
The coefficient of the identity operator is equal to
the second moment of the uniformly weighted position
grid, $\Delta x^2(N^2-1)/12$.

\paragraph{Unary}
In the unary encoding, $Q$ qubits represent $N=Q$
position-grid points within the single-excitation
subspace. We adopt Dirichlet boundary conditions,
with the position coordinates defined as in
Sec.~\ref{sec:trunc}.
The position-basis states are represented by
\begin{align}
\label{eq:unary_basis}
\ket{x_j}
=
\ket{0}^{\otimes(Q-j)}
\otimes\ket{1}
\otimes\ket{0}^{\otimes(j-1)},
\qquad j=1,2,\ldots,Q.
\end{align}
The physical Hilbert space is the single-excitation
subspace,
\begin{align}
\mathcal H_1
=
\operatorname{span}
\left\{\ket{x_j}\right\}_{j=1}^{Q},
\qquad
\dim\mathcal H_1=Q,
\end{align}
whose dimension scales linearly with the number of
qubits.

The local occupation operators and the total
excitation-number operator are defined as
\begin{align}
\label{eq:unary_num}
\hat n_j
&=\frac{I-Z_j}{2},
\qquad
Z_j=I^{\otimes(Q-j)}
\otimes Z\otimes I^{\otimes(j-1)},\\
\hat N_{\mathrm{exc}}
&=\sum_{j=1}^{Q}\hat n_j.
\end{align}
Within the single-excitation subspace $\mathcal{H}_1$, these operators
satisfy
\begin{align}
\label{eq:unary_num_properties}
\hat N_{\mathrm{exc}}\big|_{\mathcal H_1}
&=I_{\mathcal H_1},\\
\hat n_j\ket{x_k}
&=\delta_{jk}\ket{x_k},\\
\hat n_i\hat n_j\big|_{\mathcal H_1}
&=\delta_{ij}\hat n_j\big|_{\mathcal H_1}.
\end{align}
Thus, the local occupation operators act as orthogonal
projectors onto the corresponding position-basis states
within $\mathcal H_1$.

The position operator is represented by
\begin{align}
\label{eq:unary_pos}
\hat{x}\big|_{\mathcal H_1}
&=\sum_{j=1}^{Q}
x_j\hat n_j\big|_{\mathcal H_1},
\end{align}
and its powers satisfy
\begin{align}
\label{eq:unary_x_power}
\hat{x}^p\big|_{\mathcal H_1}
&=\sum_{j=1}^{Q}
x_j^p\hat n_j\big|_{\mathcal H_1}.
\end{align}
More generally, an arbitrary position-dependent
potential can be written as
\begin{align}
\label{eq:unary_potential}
V(\hat{x})\big|_{\mathcal H_1}
=
\sum_{j=1}^{Q}
V(x_j)\hat n_j\big|_{\mathcal H_1}.
\end{align}
Therefore, the potential operator requires only
single-qubit Pauli-$Z$ rotations, regardless of the
functional form of $V(x)$. Different potentials can
be implemented using the same gate structure, with
rotation angles determined by the values $V(x_j)$
at the lattice sites.

\subsubsection{Kinetic Operator Encoding}

The kinetic operator can be represented using either a Fourier-space
approach or a finite-difference discretization. For periodic boundary
conditions, the quantum Fourier transform (QFT) provides a natural
basis in which the kinetic operator is diagonal. For Dirichlet
boundary conditions, the quantum discrete sine transform (QDST)~\cite{klappenecker2001discretecosinetransformsquantum}
can also be employed to diagonalize the discrete Laplacian.
However, the QDST requires a separate circuit construction,
which is not considered in this work.

When the wavefunction remains sufficiently localized away from
the boundaries, boundary effects can be suppressed by choosing
a sufficiently large truncation length $L$.
We therefore employ the QFT-based spectral representation for
periodic boundary conditions and the finite-difference Laplacian
for Dirichlet boundary conditions.

\paragraph{Binary Encoding}

Consider the periodic position lattice defined in
Eq.~\eqref{eq:position}, with $N=2^Q$ sites encoded
using $Q$ qubits. As defined in
Eq.~\eqref{eq:bin_pos_basis}, the position-basis states
$\ket{x_j}$ are mapped to computational basis states
$\ket{j-1}$, where $j=1,\ldots,N$.

The momentum eigenstates are constructed from the
position basis through the quantum Fourier transform
(QFT),
\begin{equation}
    \ket{p_k}
    =
    \frac{1}{\sqrt{N}}
    \sum_{j=1}^{N}
    e^{2\pi\mathrm{i}(j-1)k/N}\ket{x_j},
    \qquad k=0,1,\ldots,N-1.
    \label{eq:pk}
\end{equation}
Equivalently, the QFT is represented by the unitary
operator
\begin{equation}
    F_{\mathrm{QFT}}
    =
    \frac{1}{\sqrt{N}}
    \sum_{j=1}^{N}\sum_{k=0}^{N-1}
    e^{2\pi\mathrm{i}(j-1)k/N}
    \ket{x_j}\bra{x_{k+1}},
    \label{eq:binary_qft}
\end{equation}
which satisfies
$F_{\mathrm{QFT}}\ket{x_{k+1}}=\ket{p_k}$.

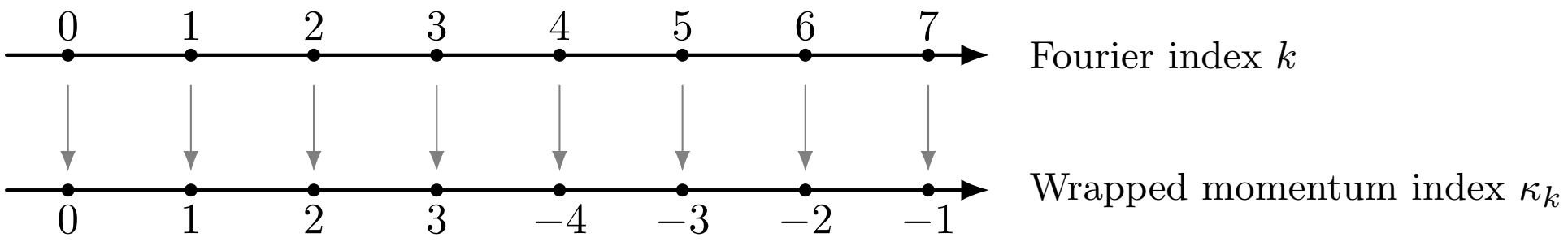
\begin{figure}[htbp]
\centering
\begin{tikzpicture}[
    >=Latex,
    line cap=round,
    line join=round,
    x=1cm,
    y=1cm
]
    \draw[->,thick] (-0.5,1.6) -- (7.5,1.6);
    \foreach \x in {0,...,7}{
        \fill (\x,1.6) circle (1.5pt);
        \node[above] at (\x,1.6) {$\x$};
    }
    \node[anchor=west,font=\small]
        at (7.7,1.6) {Fourier index $k$};

    \draw[->,thick] (-0.5,0.5) -- (7.5,0.5);
    \foreach \x/\l in {
        0/0,1/1,2/2,3/3,
        4/-4,5/-3,6/-2,7/-1
    }{
        \fill (\x,0.5) circle (1.5pt);
        \node[below] at (\x,0.5) {$\l$};
    }
    \node[anchor=west,font=\small]
        at (7.7,0.5)
        {Wrapped momentum index $\kappa_k$};

    \foreach \x in {0,...,7}{
        \draw[->,gray] (\x,1.35) -- (\x,0.65);
    }
\end{tikzpicture}
\caption{
Mapping between the Fourier index $k$ and the
wrapped momentum index $\kappa_k$ for $N=8$
with periodic boundary conditions, as defined
in Eq.~\eqref{eq:wrapped_momentum}.
}
\label{fig:momentum}
\end{figure}

The corresponding momentum eigenvalues are defined as
\begin{equation}
    p_k=\kappa_k\Delta p,
    \qquad
    \Delta p=\frac{2\pi}{N\Delta x}=\frac{\pi}{L},
    \label{eq:momentum_eigenvalues}
\end{equation}
where the wrapped momentum index is
\begin{equation}
    \kappa_k=
    \begin{cases}
        k,   & 0\leq k<N/2,\\
        k-N, & N/2\leq k<N.
    \end{cases}
    \label{eq:wrapped_momentum}
\end{equation}
This mapping, illustrated in Fig.~\ref{fig:momentum},
assigns negative momenta to Fourier indices
$k\geq N/2$. It follows from the periodicity of
discrete Fourier modes, which are defined modulo
the reciprocal lattice vector $2\pi/\Delta x$.

For the centered position coordinates in Eq.~\eqref{eq:position}, the Fourier states
in Eq.~\eqref{eq:pk} differ from the corresponding
position-space plane waves only by a
momentum-dependent phase,
\begin{equation}
    \frac{1}{\sqrt{N}}
    \sum_{j=1}^{N}
    e^{\mathrm{i}p_kx_j}\ket{x_j}
    =
    e^{-\mathrm{i}p_k(N-1)\Delta x/2}
    \ket{p_k}.
\end{equation}
This phase leaves the projectors
$\ket{p_k}\bra{p_k}$ unchanged and can therefore
be absorbed into the Fourier basis convention.

The momentum operator is diagonal in the Fourier
basis,
\begin{equation}
    \hat p
    =
    \sum_{k=0}^{N-1}
    p_k\ket{p_k}\bra{p_k},
    \label{eq:binary_momentum_operator}
\end{equation}
and the corresponding kinetic operator is
\begin{equation}
    \hat T_{\mathrm{binary}}
    =
    \frac{\hat p^{\,2}}{2m}
    =
    \sum_{k=0}^{N-1}
    \frac{p_k^2}{2m}
    \ket{p_k}\bra{p_k}.
    \label{eq:binary_kinetic_operator}
\end{equation}
The QFT diagonalizes the kinetic operator, allowing the kinetic time evolution to be implemented as a diagonal phase gate in the Fourier basis.

\paragraph{Unary Encoding}

In unary encoding, we consider $N=Q$ lattice sites
represented by $Q$ qubits, with Dirichlet boundary
conditions. The position coordinates are defined
as in Eq.~\eqref{eq:position},
\begin{equation}
    x_j=\left(j-\frac{Q+1}{2}\right)\Delta x,
    \qquad
    \Delta x=\frac{2L}{Q+1},
    \qquad j=1,\ldots,Q.
\end{equation}
The wavefunction vanishes at the two boundaries,
$\psi_0=\psi_{Q+1}=0$, corresponding to
$x_0=-L$ and $x_{Q+1}=L$.

The second-order central finite-difference
approximation to the Laplacian is
\begin{equation}
    (\nabla_D^2\psi)_j
    =
    \frac{\psi_{j+1}-2\psi_j+\psi_{j-1}}
    {\Delta x^2},
    \qquad j=1,\ldots,Q.
\end{equation}
In the position basis, the corresponding operator
is given by
\begin{align}
    \hat{\nabla}_D^2
    =&-\frac{2}{\Delta x^2}
    \sum_{j=1}^{Q}\ket{x_j}\bra{x_j}+\frac{1}{\Delta x^2}
    \sum_{j=1}^{Q-1}
    \left(
        \ket{x_{j+1}}\bra{x_j}
        +\ket{x_j}\bra{x_{j+1}}
    \right).
    \label{eq:unary_laplacian}
\end{align}

As introduced in Eq.~\eqref{eq:unary_basis},
each position-basis state corresponds to a
single excitation among the $Q$ qubits.
These states can be written as
\begin{equation}
    \ket{x_j}
    =a_j^\dagger\ket{0}^{\otimes Q},
    \qquad j=1,\ldots,Q,
\end{equation}
where $a_j^\dagger$ and $a_j$ are the
local excitation creation and annihilation
operators, respectively. They are defined by
\begin{align}
    a_j
    &=I^{\otimes(Q-j)}
    \otimes a\otimes I^{\otimes(j-1)},
    \notag\\
    a_j^\dagger
    &=I^{\otimes(Q-j)}
    \otimes a^\dagger\otimes I^{\otimes(j-1)},
\end{align}
with
$a^\dagger=\ket{1}\bra{0}$ and
$a=\ket{0}\bra{1}$.

Within the single-excitation subspace
$\mathcal H_1$, the position-space
transition operators satisfy
\begin{equation}
    \left.\ket{x_j}\bra{x_k}\;\right|_{\mathcal H_1}
    =
    \left.a_j^\dagger a_k\right|_{\mathcal H_1}.
    \label{eq:unary_transition}
\end{equation}

The local ladder operators can be expressed
in terms of Pauli operators as
\begin{equation}
    a_j^\dagger=\frac{X_j-\mathrm{i}Y_j}{2},
    \qquad
    a_j=\frac{X_j+\mathrm{i}Y_j}{2},
    \qquad
    \hat n_j=a_j^\dagger a_j
    =\frac{I-Z_j}{2}.
\end{equation}
Consequently, the nearest-neighbor hopping
terms become
\begin{equation}
    a_{j+1}^\dagger a_j
    +a_j^\dagger a_{j+1}
    =
    \frac{1}{2}
    \left(
        X_jX_{j+1}+Y_jY_{j+1}
    \right).
    \label{eq:unary_hopping}
\end{equation}
This is an exact identity for the qubit
ladder operators, which commute on different
sites.

Substituting these expressions into
Eq.~\eqref{eq:unary_laplacian}, the kinetic
operator $\hat T=-\hat{\nabla}_D^2/(2m)$
is represented within $\mathcal H_1$ by
the following excitation-number-conserving
qubit operator:
\begin{equation}
    \boxed{
    \hat T_{\mathrm{unary}}
    =
    \frac{1}{m\Delta x^2}\hat N_{\mathrm{exc}}
    -
    \frac{1}{4m\Delta x^2}
    \sum_{j=1}^{Q-1}
    \left(
        X_jX_{j+1}+Y_jY_{j+1}
    \right),
    }
    \label{eq:unary_kinetic}
\end{equation}
where
$\hat N_{\mathrm{exc}}=\sum_{j=1}^{Q}\hat n_j$
is the total excitation-number operator.

Since
$\hat N_{\mathrm{exc}}|_{\mathcal H_1}
=I_{\mathcal H_1}$,
the diagonal term contributes only a
constant energy shift within the physical
subspace. Apart from this constant term,
the unary kinetic operator consists of
nearest-neighbor $XX+YY$ hopping interactions.

Furthermore, the kinetic operator conserves
the total excitation number,
\begin{equation}
    [\hat T_{\mathrm{unary}},
    \hat N_{\mathrm{exc}}]=0.
\end{equation}
Since the potential operator is also diagonal
in the occupation-number basis, the full
Hamiltonian preserves the single-excitation
subspace under ideal time evolution.
This conservation law enables
post-selection-based error detection by
discarding measurement outcomes with
an excitation number different from one.

\subsection{Hamiltonian and time evolution}
Quantum time evolution can be implemented using product formulas,
including Suzuki--Trotter decomposition and qDRIFT~\cite{Campbell_2019}.
Other approaches include variational quantum time evolution
(VQTE)~\cite{Li_2017} and direct synthesis of the time-evolution
operator. Product formulas are broadly applicable
when the Hamiltonian can be decomposed into implementable terms.
However, repeated evolution steps can require many two-qubit gates,
making the resulting circuits susceptible to errors on
noisy intermediate-scale quantum (NISQ) devices.
VQTE can maintain a fixed circuit depth by evolving the parameters
of a chosen ansatz, but its accuracy depends strongly on that ansatz.
Direct synthesis of the exact propagator provides another approach,
although constructing and synthesizing the operator becomes
increasingly costly as the Hilbert-space dimension grows.
Here, we use second-order product formulas for both encodings. For unary encoding, the product-formula circuit is reduced to a circuit of fixed size using
local identities of the exchange gates, which are exact because the
unary dynamics of a single degree of freedom is that of free fermions
(Sec.~\ref{sec:unary_compression}).

\subsubsection{Binary Time Evolution}

In binary encoding, the kinetic operator is diagonal
in the Fourier basis, while the potential operator is
diagonal in the computational basis.
We define the diagonal kinetic phase operator as
\begin{equation}
    D_T(\Delta t)
    =
    \sum_{k=0}^{N-1}
    e^{-\mathrm{i}p_k^2\Delta t/(2m)}
    \ket{k}\bra{k},
    \label{eq:binary_kinetic_phase}
\end{equation}
where $\ket{k}$ denotes the computational basis state
labeled by the Fourier index $k$.
Using the QFT defined in Eq.~\eqref{eq:binary_qft},
the kinetic evolution is expressed as
\begin{equation}
    e^{-\mathrm{i}\hat T_{\mathrm{binary}}\Delta t}
    =
    F_{\mathrm{QFT}}
    D_T(\Delta t)
    F_{\mathrm{QFT}}^\dagger.
\end{equation}

The first- and second-order product formulas are
then given by
\begin{align}
    U_1(\Delta t)
    &=
    F_{\mathrm{QFT}}
    D_T(\Delta t)
    F_{\mathrm{QFT}}^\dagger
    e^{-\mathrm{i}V(\hat x)\Delta t},
    \label{eq:binary_trotter_first}
    \\
    U_2(\Delta t)
    &=
    e^{-\mathrm{i}V(\hat x)\Delta t/2}
    F_{\mathrm{QFT}}
    D_T(\Delta t)
    F_{\mathrm{QFT}}^\dagger
    e^{-\mathrm{i}V(\hat x)\Delta t/2}.
    \label{eq:binary_trotter_second}
\end{align}
Here, $V(\hat x)$ denotes the position-dependent
potential, so the same product-formula structure
applies to both the harmonic oscillator and
double-well systems.

For a total evolution time $t=n\Delta t$,
the time-evolution operator is approximated by
\begin{equation}
    U(t)=e^{-\mathrm{i}Ht}
    \approx    \left[U_r(\Delta t)\right]^n,
    \qquad r=1,2,
    \label{eq:binary_trotter_total}
\end{equation}
where $n$ is the number of Trotter steps.

\begin{figure*}[htbp]
    \centering
    \includegraphics[width=\textwidth]{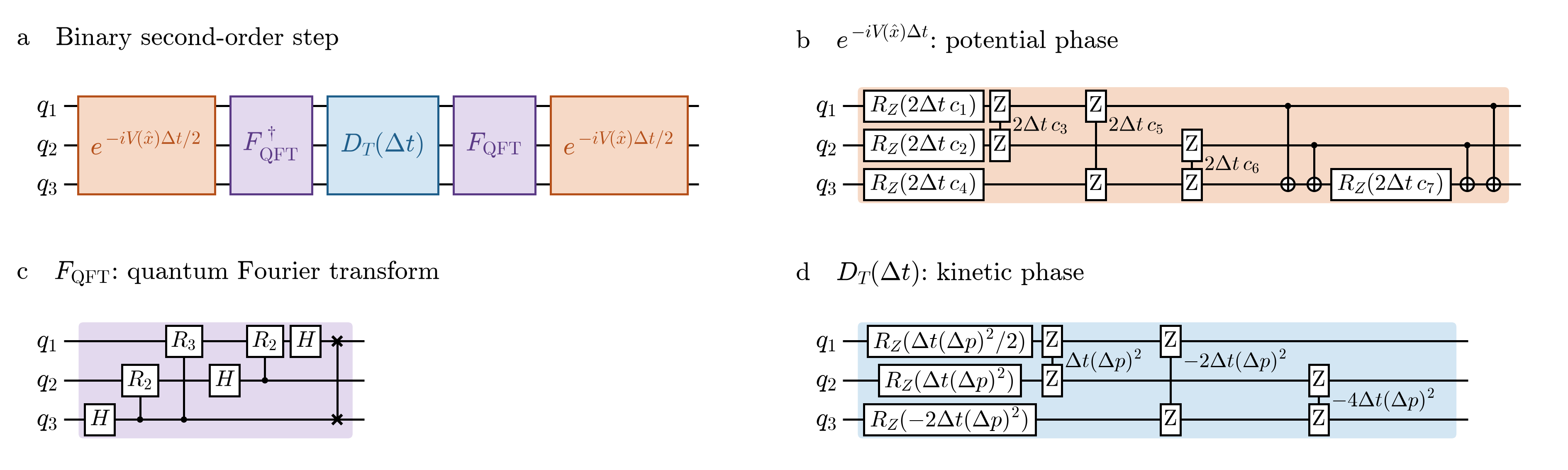}
\caption{\textbf{Logical time-evolution circuit for a harmonic oscillator using three-qubit binary encoding.}
(a) A second-order Trotter step: potential half-step, inverse QFT,
kinetic phase, QFT, and potential half-step.
(b) Diagonal potential evolution for a general potential on three qubits,
generated by Pauli-$Z$ strings with coefficients $c_k$; for the harmonic
oscillator only the two-body terms $c_5$, $c_6$ and $c_3$ are nonzero.
(c) The three-qubit QFT.
(d) Diagonal kinetic evolution in the momentum basis.}
    \label{fig:Binary_time_evolution_SHO}
\end{figure*}

Each kinetic evolution is implemented using an
inverse QFT, a diagonal kinetic phase operation,
and a QFT. In the second-order product formula,
consecutive potential half-steps from adjacent
Trotter steps can be merged according to
\begin{equation}
    e^{-\mathrm{i}V(\hat x)\Delta t/2}
    e^{-\mathrm{i}V(\hat x)\Delta t/2}
    =
    e^{-\mathrm{i}V(\hat x)\Delta t}.
\end{equation}
This reduces the number of potential evolution
blocks without changing the resulting
product-formula unitary.

Figure~\ref{fig:Binary_time_evolution_SHO} shows one step for $Q=3$. The resulting circuit consists of alternating
kinetic and potential evolution blocks, with
the kinetic terms implemented through Fourier
transforms and the potential terms implemented
through diagonal phase gates.
The number of gate operations generally increases
with the number of Trotter steps, leading to
greater circuit depth and accumulated hardware
noise.

\subsubsection{Unary Time Evolution}

In the single-excitation subspace, the unary Hamiltonian,
up to terms contributing only a global phase, is given by
\begin{equation}
  \mathcal H
  =
  -\frac{1}{4\Delta x^2}
  \sum_{j=1}^{Q-1}
  \bigl(X_jX_{j+1}+Y_jY_{j+1}\bigr)
  -\frac12
  \sum_{j=1}^{Q}V(x_j)Z_j,
  \label{eq:H-unary}
\end{equation}
where we set $m=1$.
The first term represents the kinetic Hamiltonian
$\mathcal H_K$, consisting of nearest-neighbor
$XX+YY$ interactions, while the second term
corresponds to the potential Hamiltonian
$\mathcal H_V$, represented by site-dependent
Pauli-$Z$ operators.
This Hamiltonian takes the form of an $XY$ chain
in a site-dependent field.

Both $\mathcal H_K$ and $\mathcal H_V$ conserve the total
excitation number $\hat N_{\mathrm{exc}}$ defined in
Eq.~\eqref{eq:unary_num}.
Consequently, the single-excitation subspace is
preserved under ideal time evolution.
The omitted terms contribute only a global phase
within this subspace and must be restored when
relative phases are relevant, such as in
controlled time evolution.

We approximate the time-evolution operator
using the second-order product formula,
\begin{equation}
  U(\Delta t)
  \simeq
  e^{-\mathrm{i}\mathcal H_V\Delta t/2}
  e^{-\mathrm{i}\mathcal H_K\Delta t}
  e^{-\mathrm{i}\mathcal H_V\Delta t/2}.
  \label{eq:trotter-unary}
\end{equation}

The kinetic Hamiltonian is decomposed into odd-
and even-bond contributions,
$\mathcal H_K=\mathcal H_o+\mathcal H_e$, where
\begin{align}
\mathcal H_o
&=-\frac{1}{4\Delta x^2}
\sum_{j=1,3,5,\ldots}
\left(X_jX_{j+1}+Y_jY_{j+1}\right),
\\
\mathcal H_e
&=-\frac{1}{4\Delta x^2}
\sum_{j=2,4,6,\ldots}
\left(X_jX_{j+1}+Y_jY_{j+1}\right).
\end{align}
Both sums terminate at the last allowed bond,
$j\leq Q-1$.

Within each Hamiltonian, the interactions
act on disjoint qubit pairs and therefore
commute with one another, allowing the
corresponding gates to be applied in parallel. The kinetic evolution is further approximated
using a symmetric second-order splitting,
\begin{equation}
  e^{-\mathrm{i}\mathcal H_K\Delta t}
  \simeq
  e^{-\mathrm{i}\mathcal H_o\Delta t/2}
  e^{-\mathrm{i}\mathcal H_e\Delta t}
  e^{-\mathrm{i}\mathcal H_o\Delta t/2}.
  \label{eq:unary_kinetic_trotter}
\end{equation}

\begin{figure}[htbp]
\centering
\includegraphics[width=1.0\textwidth]{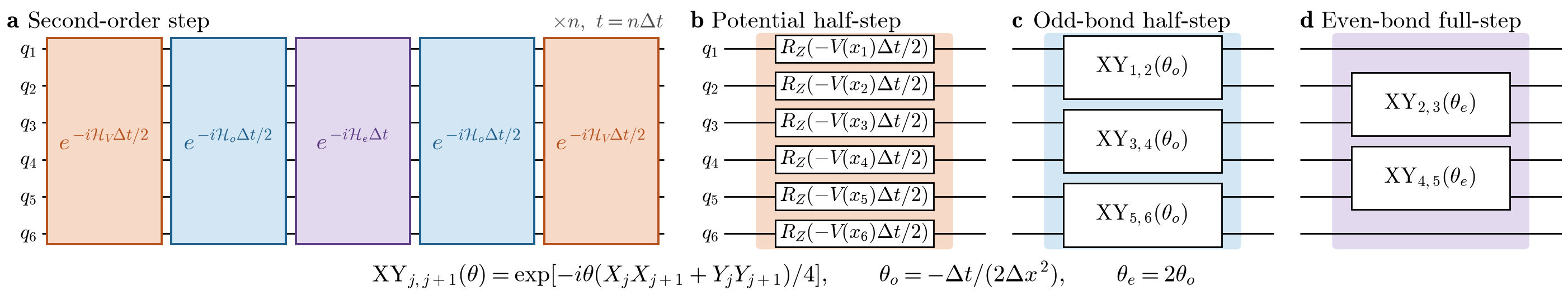}
\caption{\textbf{Logical circuit for unary time evolution on six qubits before optimization.}
(a) A single second-order Trotter step consisting of potential evolution
and kinetic evolution on even and odd bonds.
(b) Potential evolution implemented by a single $R_z$ rotation on each
qubit. The circuit structure is independent of the functional form of
the potential, which enters only through the rotation angles.
(c--d) Kinetic evolution on odd and even bonds, respectively.
Gates acting on disjoint qubit pairs within each layer can be applied
in parallel.
The circuits executed on hardware use the optimization and compilation
into native gates described in Sec.~\ref{sec:unary_compression}.}
\label{fig:unary}
\end{figure}

The exchange gate acting on neighboring qubits $(j,j+1)$
is defined as
\begin{equation}
  \mathrm{XY}_{j,j+1}(\theta)
  =
  \exp\left[
    -\frac{\mathrm{i}\theta}{4}
    \left(X_jX_{j+1}+Y_jY_{j+1}\right)
  \right],
  \label{eq:unary_xy_gate}
\end{equation}
with rotation angles
\begin{equation}
  \theta_o=-\frac{\Delta t}{2\Delta x^2},
  \qquad
  \theta_e=-\frac{\Delta t}{\Delta x^2}
\end{equation}
for the odd- and even-bond layers, respectively.
The odd-bond layers are applied before and after the
even-bond layer, realizing the symmetric kinetic decomposition.
Figure~\ref{fig:unary} shows one Trotter step before optimization.
The circuits executed on hardware are obtained from it as described
in Sec.~\ref{sec:unary_compression}.

For execution on quantum hardware, the $R_Z$
rotations and $\mathrm{XY}(\theta)$ gates are
transpiled into the native gate set of each QPU.

\subsubsection{Circuit optimization for unary evolution}
\label{sec:unary_compression}

\begin{figure}[p]
    \centering
    \includegraphics[
        width=\linewidth,
        height=0.82\textheight,
        keepaspectratio
    ]{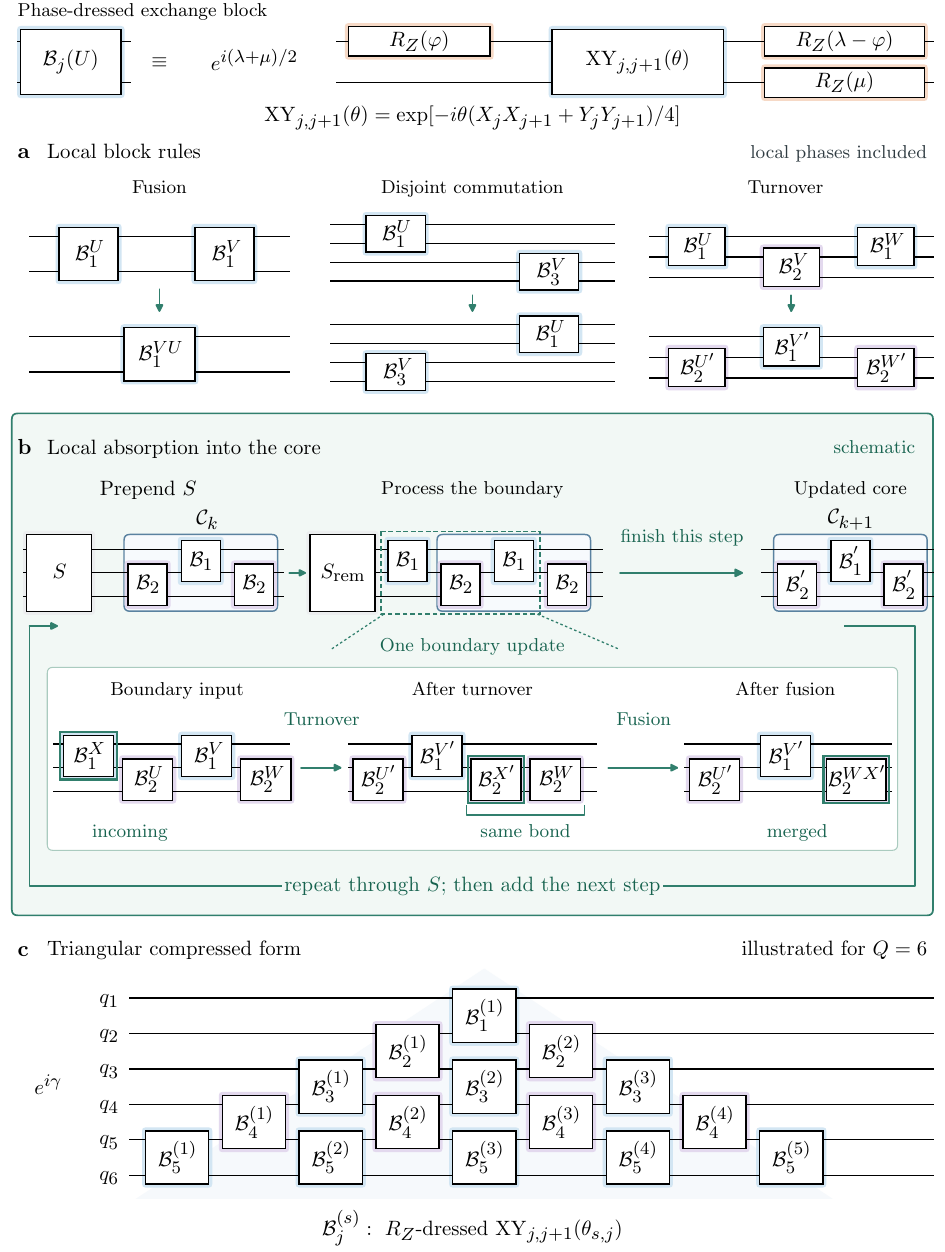}
\caption{\textbf{Optimization of unary evolution circuits.}
The top identity resolves each $\mathcal B_j(U)$ into one
$\mathrm{XY}_{j,j+1}(\theta)$ gate, local $R_Z$ rotations
and the retained scalar factor; the upper and lower wires
are $j$ and $j+1$. Superscripts on blocks label their matrices.
(a) Fusion, disjoint commutation and phase-aware turnover.
(b) An additional step enters the core; the enlargement shows
turnover followed by fusion on the same bond.
Circuits execute left to right, so $X'$ followed by $W$ fuses
to $WX'$. Updates repeat until the entire step is absorbed;
$S_{\rm rem}$ denotes its unprocessed factors.
(c) The optimized triangular circuit, illustrated for $Q=6$;
slots have independent parameters and may be identities.
Local phases are included in the blocks, while the global
scalar phase is retained separately. State preparation is
outside the displayed optimization.}
\label{fig:unary_compression}
\end{figure}

We optimize the unary evolution circuit built from the exchange gates
$\mathrm{XY}_{j,j+1}(\theta)$ and local $R_Z$ rotations.
We use the exchange-gate convention in Eq.~\eqref{eq:unary_xy_gate}, implemented as \texttt{XXPlusYYGate(theta, 0)}.
For optimization, we combine an exchange gate and local $R_Z$
rotations into a phase-dressed exchange block $\mathcal B_j(U)$,
as explicitly decomposed at the top of
Fig.~\ref{fig:unary_compression}.
Here $U\in\mathrm U(2)$ is the block's action on the ordered
one-excitation states $(|1_j0_{j+1}\rangle,|0_j1_{j+1}\rangle)$,
including their full phase information.
The scalar factor in the displayed decomposition is retained.
Setting all phase-rotation angles to zero recovers
$\mathrm{XY}_{j,j+1}(\theta)$.
The blocks are rearranged using fusion on the same bond,
commutation on disjoint bonds, and phase-aware turnover on
neighboring bonds~\cite{Camps_2022}, as illustrated in
Fig.~\ref{fig:unary_compression}(a).
Turnover replaces the bond sequence
$(j,j+1),(j+1,j+2),(j,j+1)$ with the reversed alternating
sequence, updating both exchange and phase parameters.
Repeated application of these rules absorbs $n$ Trotter steps
[Fig.~\ref{fig:unary_compression}(b)] into an identity-padded
triangular circuit [Fig.~\ref{fig:unary_compression}(c)] with
\begin{equation}
    N_{\mathrm{block}}\leq\frac{Q(Q-1)}{2},
    \qquad
    D_{\mathrm{block}}\leq 2Q-3,
    \label{eq:unary_compression_bounds}
\end{equation}
independently of $n$~\cite{Camps_2022}.
Here $D_{\mathrm{block}}$ counts logical two-qubit layers;
identity slots may be omitted.
All diagonal and scalar phases are retained, so the optimized circuit
preserves the original product-formula unitary on $\mathcal H_1$ up to
numerical roundoff and leaves its product-formula error unchanged. The
parameters are updated classically
through local $2\times2$ blocks and $3\times3$ turnover
factorizations, without constructing the full
$Q\times Q$ evolution matrix.

The optimization is exact because, within $\mathcal H_1$, the unary
Hamiltonian is quadratic: under the Jordan--Wigner transformation the
$XY$ chain in a site-dependent field maps to free fermions, and
$\mathcal H_1$ is the one-particle sector. Each block acts as a
two-level rotation of the one-particle wavefunction, and the
triangular form of Fig.~\ref{fig:unary_compression}(c) is the
Givens-rotation decomposition of a $Q\times Q$ unitary used for
linear-optical interferometers~\cite{Reck:1994ujl,Clements:2016rkw} and for free-fermion state preparation~\cite{Kivlichan:2018bqq,Jiang:2017pyp}.
Let $S\equiv S(\Delta t)$ denote the unary circuit step obtained
from Eq.~\eqref{eq:trotter-unary} using the kinetic splitting
in Eq.~\eqref{eq:unary_kinetic_trotter}.
The optimized circuit implements $S^n$ on $\mathcal H_1$,
including any scalar phase supplied to the optimization,
up to numerical roundoff. At fixed $Q$, its size bound is
independent of $n$, whereas its rotation angles depend on
$\Delta t$ and $n$. 
Noiseless product-formula baselines for both encodings are presented
in Appendix~\ref{app:compilation}; the unary baseline is evaluated
without the fixed-size optimization.

In the adopted number-conserving synthesis, each block is
locally equivalent to an exchange gate and requires at most
two CNOTs (or CZs with local rotations)~\cite{Zhang_2024}.
Two variable-angle Ising
entanglers with local basis rotations also suffice when
available. Thus the optimized evolution requires at most
$Q(Q-1)$ entangling gates before routing.
When implemented separately, the state-preparation chain of
Sec.~\ref{sec:prep} adds at most $2(Q-1)$ entangling gates.
For the IBM double-well circuits, preparation is separate from the
optimized evolution. In the IonQ double-well circuits, the
number-conserving preparation rotations are instead merged with
the evolution during optimization, while the initial $X$ seed remains
outside the optimized circuit. Table~\ref{tab:unary_gate_costs}
reports resources for the respective implementations.
Since the logical gates conserve excitation number,
post-selection on $\mathcal H_1$ remains applicable.

Algorithm~\ref{alg:unary_compression_simple} uses left-boundary
insertion: the factors of $S$ are therefore processed in
reverse circuit order. Local phase information stays in
the blocks, while the scalar phase is accumulated separately.

\begin{algorithm}[H]
\caption{Unary time-evolution circuit optimization}
\label{alg:unary_compression_simple}
\begin{algorithmic}[1]
\Require Trotter step $S$, number of steps $n\geq0$,
         number of qubits $Q\geq2$
\Ensure Circuit implementing $S^n$ on $\mathcal H_1$
\State $(\mathcal L_S,\gamma_S)\gets\textsc{ExtractBlocks}(S)$
\Statex \hspace{\algorithmicindent}$\mathcal L_S$: circuit order;
         $\gamma_S$: step scalar phase
\State $T\gets$ identity-padded triangle on $Q$ qubits;
       $\gamma\gets0$
\For{$k=1,\ldots,n$}
    \ForAll{$B\in\operatorname{reverse}(\mathcal L_S)$}
        \State Prepend and absorb $B$ into $T$ by commutation,
               turnover, and fusion
        \State Update block parameters and retain all local phases
    \EndFor
    \State $\gamma\gets\gamma+\gamma_S$
\EndFor
\State $(C,\delta\gamma)\gets\textsc{EmitAndMerge}(T)$
       \Comment{$T=e^{i\delta\gamma}C$}
\Statex \hspace{\algorithmicindent}Emit $\mathrm{XY}$ gates and merge
         residual $R_Z$ rotations
\State Attach global phase $\gamma+\delta\gamma$ to $C$
\State \Return $C$
\end{algorithmic}
\end{algorithm}

\FloatBarrier

\subsection{Preparation of initial states}\label{sec:prep}

We describe circuit preparation of the grid states specified
by Eqs.~\eqref{eq:ho_initial} and~\eqref{eq:dw_initial}.
Model-specific parameters are given in
Sec.~\ref{sec:models}.
Here, $\psi(x_j)$ denotes a normalized grid amplitude,
with $\sum_j|\psi(x_j)|^2=1$.

\subsubsection{Binary encoding}

For binary encoding, the target amplitudes are supplied to
\texttt{Qiskit}'s \texttt{StatePreparation} routine in the
position-basis order used above. The routine acts on the binary
register and is decomposed into the processor's supported gates.
Its preparation cost is included in the signal-circuit resources.

\subsubsection{Unary encoding}

For unary encoding, only the $Q$ amplitudes in the
single-excitation subspace are required. We first excite the
qubit associated with the first site. With the tensor ordering
of Eq.~\eqref{eq:unary_basis},
\begin{equation}
    \ket{\psi^{(1)}}=X_1\ket{0}^{\otimes Q}
    =\ket{0}^{\otimes(Q-1)}\otimes\ket{1}
    =\ket{x_1}.
    \label{eq:prep_start}
\end{equation}
Subsequent rotations transfer amplitude between adjacent sites
while remaining in the single-excitation subspace.

A real amplitude-transfer gate can be written in terms of the
exchange gate in Eq.~\eqref{eq:unary_xy_gate} as
\begin{equation}
    G_{j,j+1}(\theta)
    =R_{Z,j}\!\left(-\frac{\pi}{2}\right)
    \mathrm{XY}_{j,j+1}(2\theta)
    R_{Z,j}\!\left(\frac{\pi}{2}\right).
    \label{eq:Gprep}
\end{equation}
Its action in the one-excitation sector is
\begin{align}
    G_{j,j+1}(\theta)\ket{x_j}
        &=\cos\theta\ket{x_j}+\sin\theta\ket{x_{j+1}},\notag\\
    G_{j,j+1}(\theta)\ket{x_{j+1}}
        &=-\sin\theta\ket{x_j}+\cos\theta\ket{x_{j+1}}.
    \label{eq:Gprep_action}
\end{align}
The factor $2\theta$ in Eq.~\eqref{eq:Gprep} follows from the
exchange-gate angle convention used for time evolution.

To distribute the amplitude magnitudes, define
\begin{equation}
    r_j=\sqrt{\sum_{k=j}^{Q}|\psi(x_k)|^2}.
    \label{eq:prep_tail}
\end{equation}
After the first $j-1$ rotations, the state is
\begin{equation}
    \ket{\psi^{(j)}}
    =\sum_{k=1}^{j-1}|\psi(x_k)|\ket{x_k}+r_j\ket{x_j}.
    \label{eq:r_angle}
\end{equation}
The next angle is chosen as
\begin{equation}
    \theta_j=
    \begin{cases}
        \operatorname{atan2}\!\left(r_{j+1},|\psi(x_j)|\right),
            & r_j>0,\\
        0, & r_j=0,
    \end{cases}
    \qquad j=1,\ldots,Q-1.
    \label{eq:theta_prep}
\end{equation}
This gives $r_j\cos\theta_j=|\psi(x_j)|$ and
$r_j\sin\theta_j=r_{j+1}$ without dividing by a vanishing
tail norm. Once $r_j=0$, all remaining rotations can be omitted.

Local phase gates then restore the phases of the target
amplitudes. Writing $\varphi_j=\arg\psi(x_j)$, with
$\varphi_j=0$ for a zero amplitude, the complete preparation is
\begin{equation}
    \ket{\psi}
    =\left[\prod_{j=1}^{Q}e^{\mathrm{i}\varphi_j\hat n_j}\right]
    G_{Q-1,Q}(\theta_{Q-1})\cdots
    G_{2,3}(\theta_2)G_{1,2}(\theta_1)\ket{x_1}.
    \label{eq:unary_preparation}
\end{equation}
Here $\hat n_j$ is the occupation operator in
Eq.~\eqref{eq:unary_num}, and $G_{1,2}$ acts first.
For a real target, the phase gates
reduce to sign corrections, including a possible negative final
amplitude.

\begin{figure}[htbp]
    \centering
    \includegraphics[width=\linewidth]{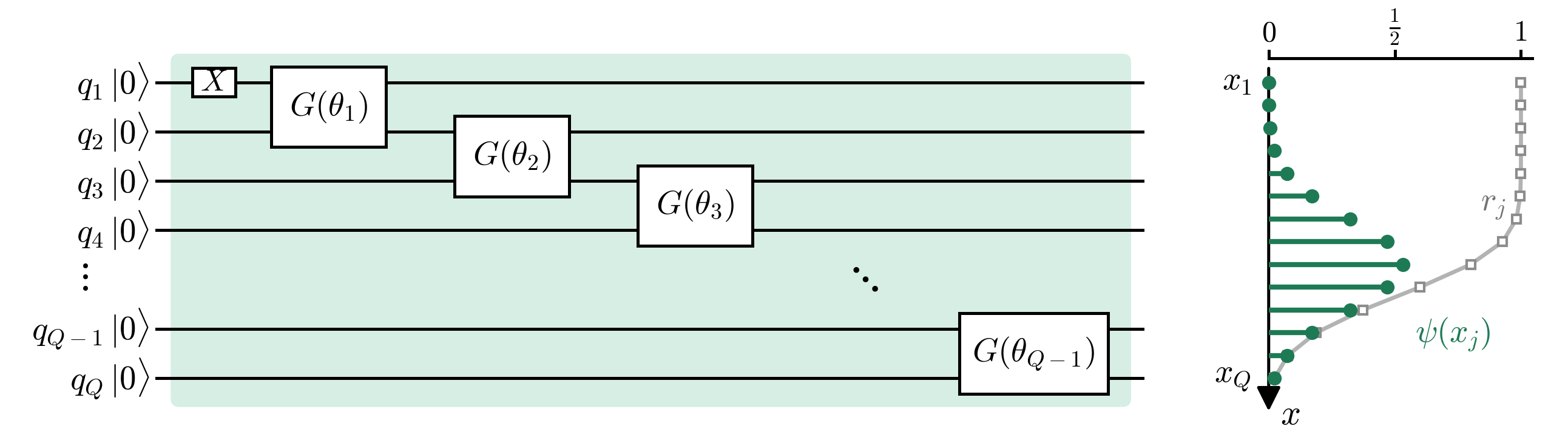}
\caption{\textbf{Schematic circuit for unary initial-state preparation.}
An $X$ gate creates a single excitation on the first qubit.
Sequential $G_{j,j+1}(\theta_j)$ rotations then distribute the excitation
along the register, setting the target amplitude magnitudes.
The green profile shows the amplitudes already assigned, while the
gray profile represents the remaining norm $r_j$.}
    \label{fig:unary_prep}
\end{figure}

The preparation requires at most $Q-1$ nearest-neighbor amplitude-transfer gates $G_{j,j+1}$ and local phase rotations (Fig.~\ref{fig:unary_prep}). For the coupled-oscillator benchmarks in Sec.~\ref{sec:coupled}, the product initial state is prepared separately in each register.

\FloatBarrier

\section{Hardware results}\label{sec:models}

We examine the harmonic oscillator, symmetric double well and
two coupled-oscillator models on IBM Phoenix and IonQ Forte.
Measured spatial distributions and moments are compared with
exact finite-grid evolution and the intended noiseless product
formula. Continuum references provide an additional physical
comparison, with their construction specified for each model.
The DW observation windows cover the first-transfer neighborhood,
extending to $t=26.4$ for binary encoding and $t=30.4$ for unary encoding.

Measurements with matching physical conditions are pooled at
the distribution level before moments are evaluated. IBM weights
are integer counts, whereas IonQ weights are returned probabilities
multiplied by the reported shots. For unary encoding, outcomes are
accepted only when the measured excitation number defined in Eq.~\eqref{eq:unary_num} equals one in each register.
Writing the accepted site weights as $w_j$, we calculate
\begin{equation}
\hat P_j=\frac{w_j}{\sum_k w_k},\qquad
\langle x\rangle=\sum_jx_j\hat P_j,\qquad
\operatorname{Var}(x)=\sum_jx_j^2\hat P_j-\langle x\rangle^2.
\label{eq:review_measurement}
\end{equation}
For two coordinates, joint weights $w_{ij}$ are normalized
by $\sum_{ij}w_{ij}$, and coordinate moments are obtained from
the corresponding marginals. Unary results are conditional
on this selection, which rejects measured leakage but does
not remove errors within the encoding space. The IonQ experiments reported here were performed with
debiasing disabled.

\begin{table*}[!htbp]
\centering
\caption{Discretization parameters for the QPU simulations.
Binary encoding uses periodic boundary conditions (PBC),
while unary encoding uses Dirichlet boundary conditions (DBC).
$N$ and $Q$ denote the numbers of lattice sites and qubits,
respectively.}
\label{tab:benchmark_grids}
\small
\renewcommand{\arraystretch}{1.15}
\begin{tabular*}{\textwidth}{@{\extracolsep{\fill}}lllccccc@{}}
\toprule
Model & Processor & Encoding & $N$ & $Q$
& $\Delta x$ & $\Delta t$ & Boundary \\
\midrule
SHO & IBM Phoenix & Binary & 8 & 3 & 0.710 & 0.400 & PBC \\
SHO & IonQ Forte & Binary & 8 & 3 & 0.710 & 0.400 & PBC \\
SHO & IBM Phoenix & Unary & 8 & 8 & 0.710 & 0.100 & DBC \\
SHO & IonQ Forte & Unary & 8 & 8 & 0.710 & 0.100 & DBC \\
\midrule
DW & IBM Phoenix & Binary & 16 & 4 & 0.3748 & 0.200 & PBC \\
DW & IonQ Forte & Binary & 16 & 4 & 0.3748 & 0.200 & PBC \\
DW & IBM Phoenix & Unary & 14 & 14 & 0.3748 & 0.013793 & DBC \\
DW & IonQ Forte & Unary & 14 & 14 & 0.3748 & 0.013793 & DBC \\
\bottomrule
\end{tabular*}
\end{table*}

On IBM Phoenix, several copies of each circuit, referred to as regions,
were executed in parallel on disjoint sets of qubits within the same job. The raw shot budgets, pooled regional observation counts and unary
post-selection retention fractions are summarized in
Appendix~\ref{app:regional_shot_budget}. Simultaneously measured
regions share the same joint shots, so pooling their observations
does not create independent circuit repetitions.

Error bars show nominal central 95\% bootstrap percentile intervals
obtained separately at each measurement time from 2,000 resamples.
For IBM data, whole joint-shot records are resampled while keeping
simultaneous regional outcomes aligned. For IonQ data, multinomial
resamples use the returned probabilities and reported shot budgets.
Unary post-selection and normalization are reapplied within each
resample.

\subsection{Harmonic oscillator}\label{sec:ho}

We use the harmonic potential in Eq.~\eqref{eq:potentials}
and the Gaussian initial state in Eq.~\eqref{eq:ho_initial}.
Their parameters are listed in Table~\ref{tab:sho_initial_parameters},
with the grids specified in Table~\ref{tab:benchmark_grids}.
The continuum curve follows Eq.~\eqref{eq:ho-moments} with the Gaussian center $x_0\approx 0.365$ and $s^2=5/3$, rather than the sampled grid mean. Its constant variance is $5/6$, whereas the normalized sampled grid has the moments in Table~\ref{tab:sho_initial_parameters}.

\begin{table*}[h]
\centering
\caption{SHO potential and initial-state parameters
used for both encodings and processors.
The initial Gaussian amplitudes are
$\psi_j(0)\propto
\exp[-(x_j-x_0)^2/(2s^2)]$.}
\label{tab:sho_initial_parameters}
\small
\renewcommand{\arraystretch}{1.15}
\begin{tabular*}{\textwidth}{
@{}l@{\hspace{1.5em}}r
@{\extracolsep{\fill}}
l@{\hspace{1.5em}}r@{}
}
\toprule
Quantity & Value & Quantity & Value \\
\midrule
Initial state & Gaussian
& Gaussian width $s^2$ & 1.667 \\
$k$ & 0.360
& Initial mean $\langle x\rangle(0)$ & 0.358 \\
$\omega$ & 0.600
& Initial variance $\operatorname{Var}(x)(0)$ & 0.811 \\
Gaussian center $x_0$ & 0.365
& & \\
\bottomrule
\end{tabular*}
\end{table*}

\begin{figure}[htbp]
    \centering
    \includegraphics[
        width=0.88\linewidth,
        height=0.36\textheight,
        keepaspectratio]{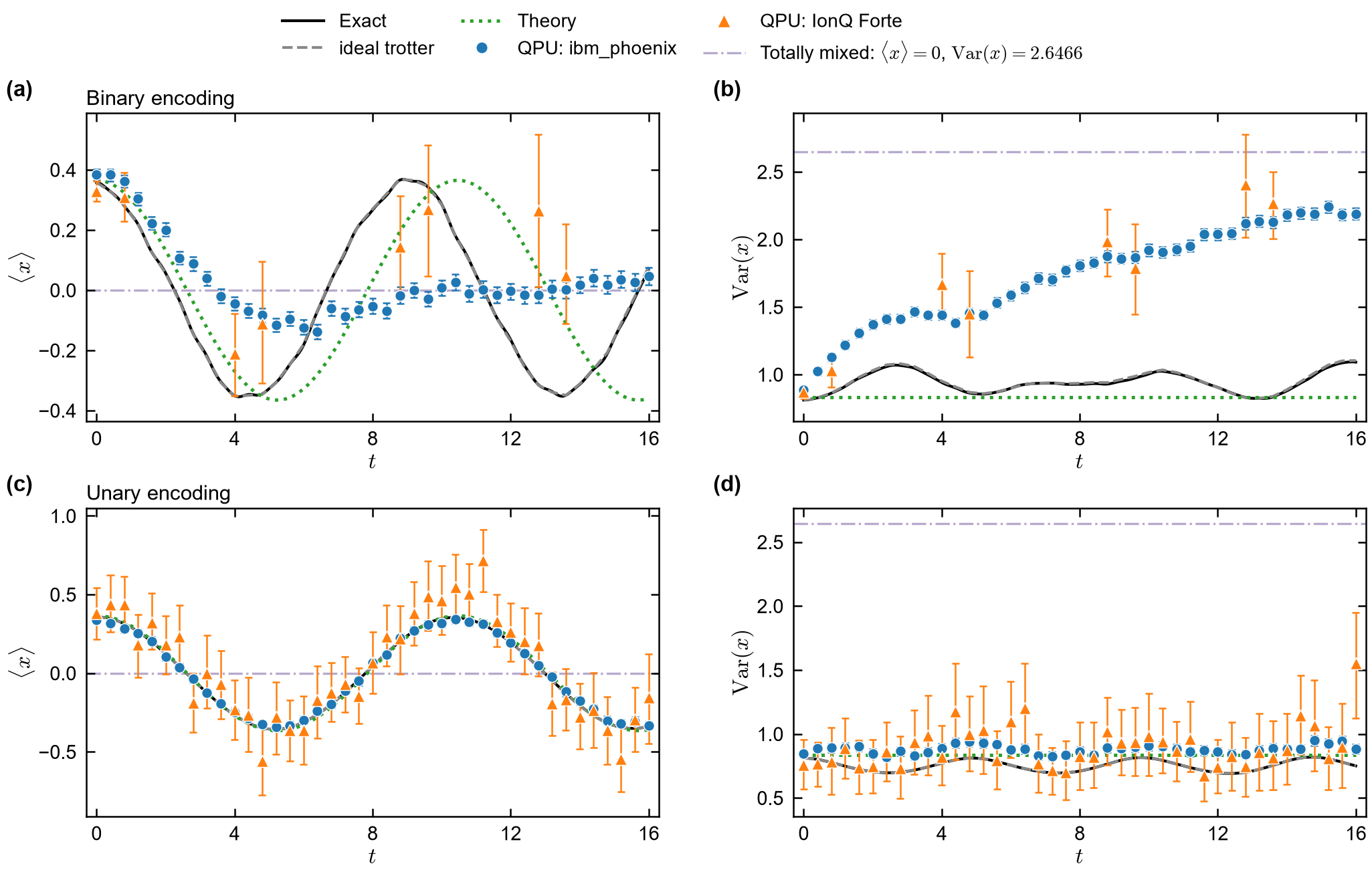}
    \caption{\textbf{Simple harmonic oscillator benchmark on an eight-point grid.} Panels (a,b) and (c,d) show binary and unary results, respectively, with the position mean (left) and variance (right) plotted versus time. Blue circles and orange triangles denote IBM Phoenix and IonQ Forte data. Black solid and gray dashed curves show the encoding-specific exact-grid and ideal product-formula predictions, respectively. Green dotted curves show the continuum predictions of Eq.~\eqref{eq:ho-moments} for an initial Gaussian centered at $x_0\simeq0.365$. Purple lines indicate the maximally mixed reference values, $\braket{x}=0$ and $\mathrm{Var}(x)=2.6466$. Unary results extend to $t=16$ for both devices, while the IonQ binary results end at $t=13.6$.}
    \label{fig:sho_moments}
\end{figure}

Figure~\ref{fig:sho_moments} compares the coordinate means and
variances for both encodings and processors. IBM observations span 41 times over $0\le t\le16$, and
IonQ unary observations cover the same times.  For IonQ binary
encoding, a budget-constrained measurement schedule samples
eight selected times over $0\le t\le13.6$, prioritizing times
with relatively large predicted mean-position signals. In the binary data, the mean trajectories depart
substantially from their references and the distributions broaden.
The unary mean positions follow the predicted oscillatory trajectories more closely, while the measured variances still show offsets and
fluctuations relative to the finite-grid predictions.

The constant continuum variance and the oscillating finite-grid
variance illustrate why these references must be distinguished.
These observations characterize the selected
implementations and post-processing: the differing finite
Hamiltonians, circuit resources, shot budgets and acquisitions do
not isolate an effect of the encoding alone. 

\subsection{Double-well potential}\label{sec:dw}

We use the double-well potential in Eq.~\eqref{eq:potentials}
and the right-localized initial state in Eq.~\eqref{eq:dw_initial}.
Table~\ref{tab:dw_initial_parameters} lists the parameters and
lowest-doublet energies for the grids in
Table~\ref{tab:benchmark_grids}.

Both encodings use the same potential parameters and grid spacing $\Delta x=0.3748$, but different time steps: $\Delta t=0.2$ for binary and $\Delta t  \simeq0.01379$ for unary. Each initial state is constructed from the lowest doublet
of its corresponding finite-grid Hamiltonian. The binary
implementation uses QFT-based kinetic evolution with periodic
boundary conditions, whereas the unary implementation uses
a three-point finite-difference kinetic operator with Dirichlet
boundary conditions. Their predicted transfer times are
$t_{\mathrm{tun}}=26.655$ and $28.088$, respectively. The continuum reference of Table~\ref{tab:dw_initial_parameters} has $\Delta E=0.1175$ and $t_{\mathrm{tun}}=26.75$. The binary and unary splittings differ from that reference by approximately $+0.34\%$ and $-4.78\%$. These are total differences between specified finite models; they do not separately identify finite-box and kinetic-stencil contributions.

The binary records comprise 12 IBM and eight IonQ times through $t=26.4$. The unary records comprise 74 IBM and 22 IonQ times through $t=30.4$; the latter has 192 reported shots per time, including 21 times through $t=28.8$. These windows include the first-transfer neighborhood rather than a full return. Unary statistics are conditioned on exactly one excitation; binary statistics use all decoded positions.

\begin{table*}[t]
\centering
\caption{\textbf{Double-well parameters and lowest energy doublets.}
The binary and unary discretizations share the same potential parameters,
$a=1.5$ and $V_0=2.901$.
The continuum reference is obtained by diagonalizing the Hamiltonian
on a fine grid of $N=3001$ interior points over $x\in[-6,6]$.
The energy splitting is $\Delta E=E_1-E_0$, with the corresponding
tunneling time $t_{\mathrm{tun}}=\pi/\Delta E$.}
\label{tab:dw_initial_parameters}
\small
\renewcommand{\arraystretch}{1.15}
\begin{tabular*}{\textwidth}{@{\extracolsep{\fill}}lccc@{}}
\toprule
Quantity & Continuum reference & Binary & Unary \\
\midrule
$a$ & \multicolumn{3}{c}{1.500} \\
$V_0$ & \multicolumn{3}{c}{2.901} \\
$N$ & 3001 & 16 & 14 \\
$\Delta x$ & 0.003997 & 0.3748 & 0.3748 \\
$E_0$ & 1.410 & 1.410 & 1.380 \\
$E_1$ & 1.528 & 1.528 & 1.492 \\
$\Delta E$ & 0.1175 & 0.1179 & 0.1118 \\
$t_{\mathrm{tun}}$ & 26.75 & 26.65 & 28.09 \\
\bottomrule
\end{tabular*}
\end{table*}

\begin{figure}[!tp]
    \centering
    \includegraphics[
        width=0.95\linewidth,
        height=0.68\textheight,
        keepaspectratio
    ]{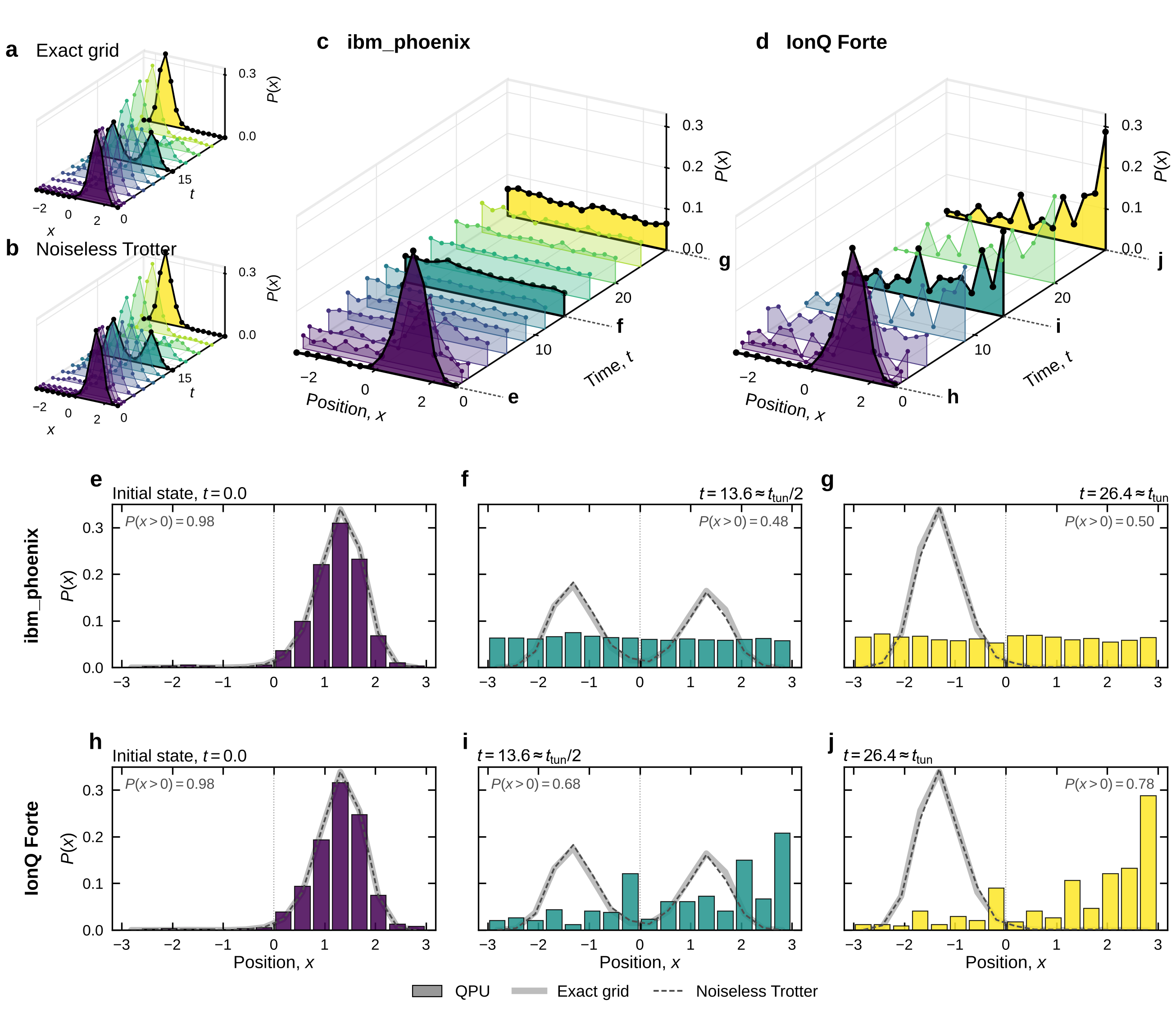}
 \caption{\textbf{Position-space probability distributions for double-well dynamics with 16-site binary encoding.} Waterfall plots show (a) exact-grid, (b) ideal product-formula, (c) IBM Phoenix, and (d) IonQ Forte evolution during the first transfer. Snapshots at $t=0,\,13.6,\,26.4$ are shown for IBM Phoenix (e--g) and IonQ Forte (h--j). Colored bars represent measured site probabilities, while gray filled profiles and gray dashed curves show the exact-grid and ideal product-formula predictions, respectively. Color encodes time, and $P(x>0)$ denotes the right-well probability.}
    \label{fig:binary_dw_overview3d_first_transfer}
\end{figure}

\begin{figure}[htbp]
    \centering
    \includegraphics[
        width=0.88\linewidth,
        height=0.36\textheight,
        keepaspectratio]{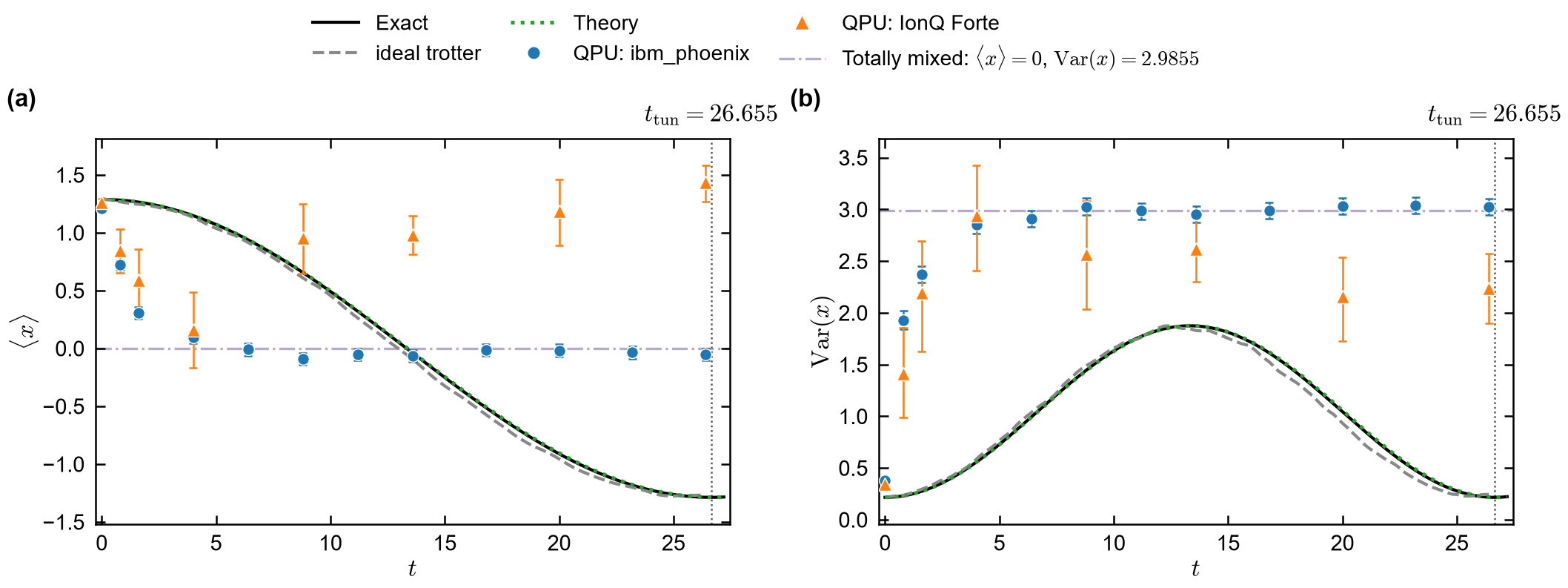}
\caption{\textbf{Double-well dynamics with 16-site binary encoding.}
(a,b) Position mean and variance, respectively, up to $t=26.4$
with a Trotter step size of $\Delta t=0.2$.
Blue circles and orange triangles denote IBM Phoenix and IonQ Forte
measurements, respectively.
Black solid, gray dashed, and green dotted curves represent the
exact-grid, ideal product-formula, and numerical reference predictions.
Purple horizontal lines indicate the maximally mixed reference values,
$\braket{x}=0$ and $\mathrm{Var}(x)=2.9855$.
The vertical dotted line marks the binary-grid tunneling time,
$t_{\mathrm{tun}}=26.655$.}
    \label{fig:dw_binary_first_transfer}
\end{figure}

Figure~\ref{fig:dw_binary_first_transfer} contains two panels, with both processors shown in each: mean position and variance. IBM's late mean lies near zero and its variance near the maximally mixed state value 2.9855; the expected right-to-left transfer is not reproduced. IonQ also does not reproduce that transfer, retaining a positive mean and a right-well preference near the endpoint. The site distributions in Fig.~\ref{fig:binary_dw_overview3d_first_transfer} show the same behavior: on IBM Phoenix the distribution is nearly uniform over the grid at $t=13.6$ and $26.4$, with $P(x>0)=0.48$ and $0.50$, whereas on IonQ Forte most of the weight remains in the right well, with $P(x>0)=0.68$ and $0.78$.

\begin{figure}[htbp]
    \centering
    \includegraphics[
        width=0.88\linewidth,
        height=0.36\textheight,
        keepaspectratio
    ]{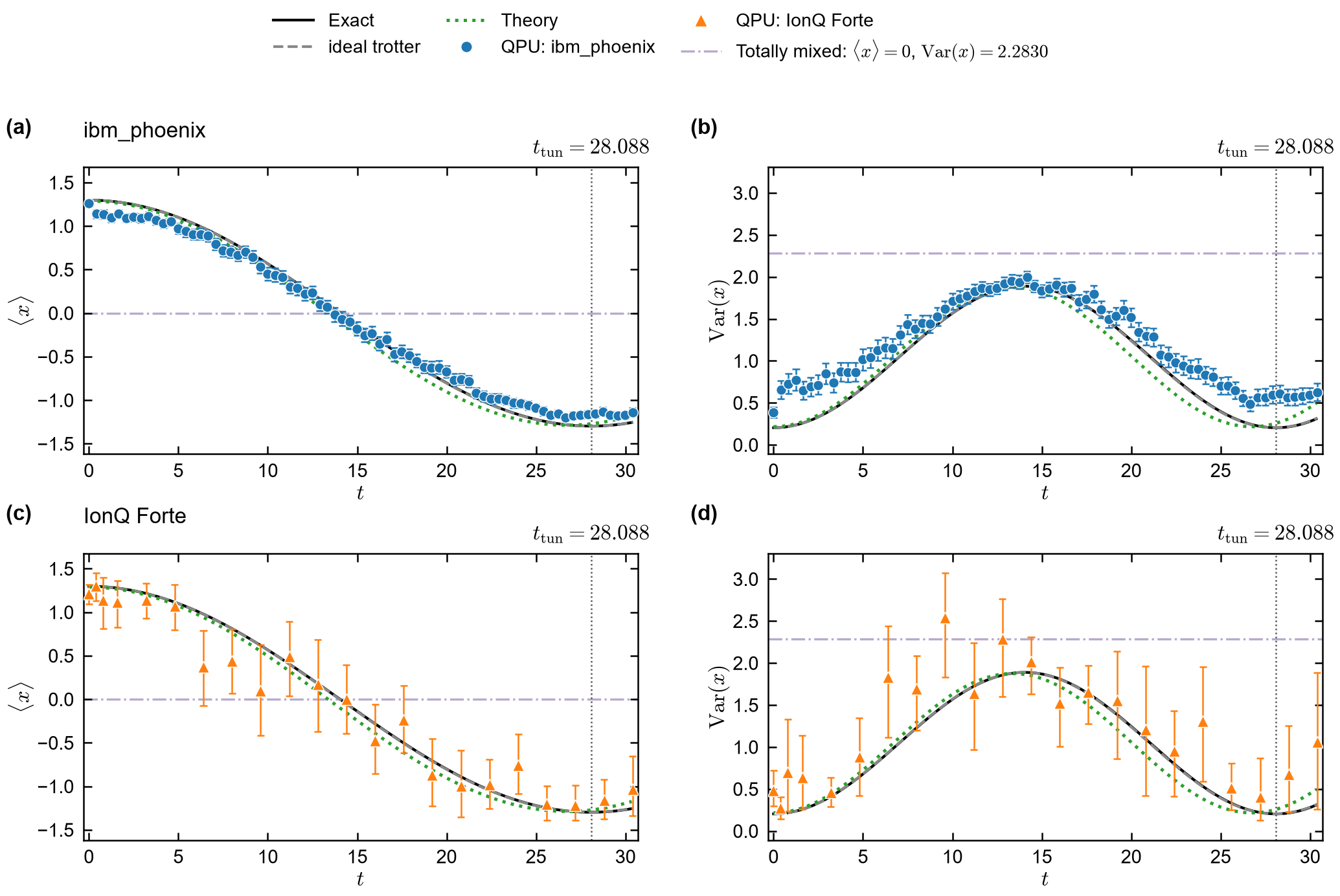}
\caption{\textbf{Double-well dynamics with 14-site unary encoding.}
(a,b) IBM Phoenix and (c,d) IonQ Forte results for the position
mean (left) and variance (right), up to $t=30.4$ with a Trotter
step size of $\Delta t\simeq0.01379$.
Blue circles and orange triangles denote IBM Phoenix and IonQ Forte
measurements, respectively.
Black solid, gray dashed, and green dotted curves represent the
exact-grid, ideal product-formula, and numerical reference predictions.
Purple horizontal lines indicate the maximally mixed reference values,
$\braket{x}=0$ and $\mathrm{Var}(x)=2.2830$.
The vertical dotted line marks the unary-grid tunneling time,
$t_{\mathrm{tun}}=28.088$.}
    \label{fig:dw_first_transfer}
\end{figure}

Figure~\ref{fig:dw_first_transfer} compares the conditional unary moments with the same 14-site reference model on both processors. IBM resolves the change from positive to negative mean position, while its variance remains above the grid reference near the ends of the window. IonQ also resolves transfer, with larger fluctuations in its mean and width. Intermediate broadening followed by narrowing is consistent with redistribution between the wells, but the residual width discrepancies show that agreement of the mean alone is insufficient to establish agreement of the full position distribution. 

\begin{figure}[!tp]
    \centering
    \includegraphics[
        width=0.95\linewidth,
        height=0.68\textheight,
        keepaspectratio
    ]{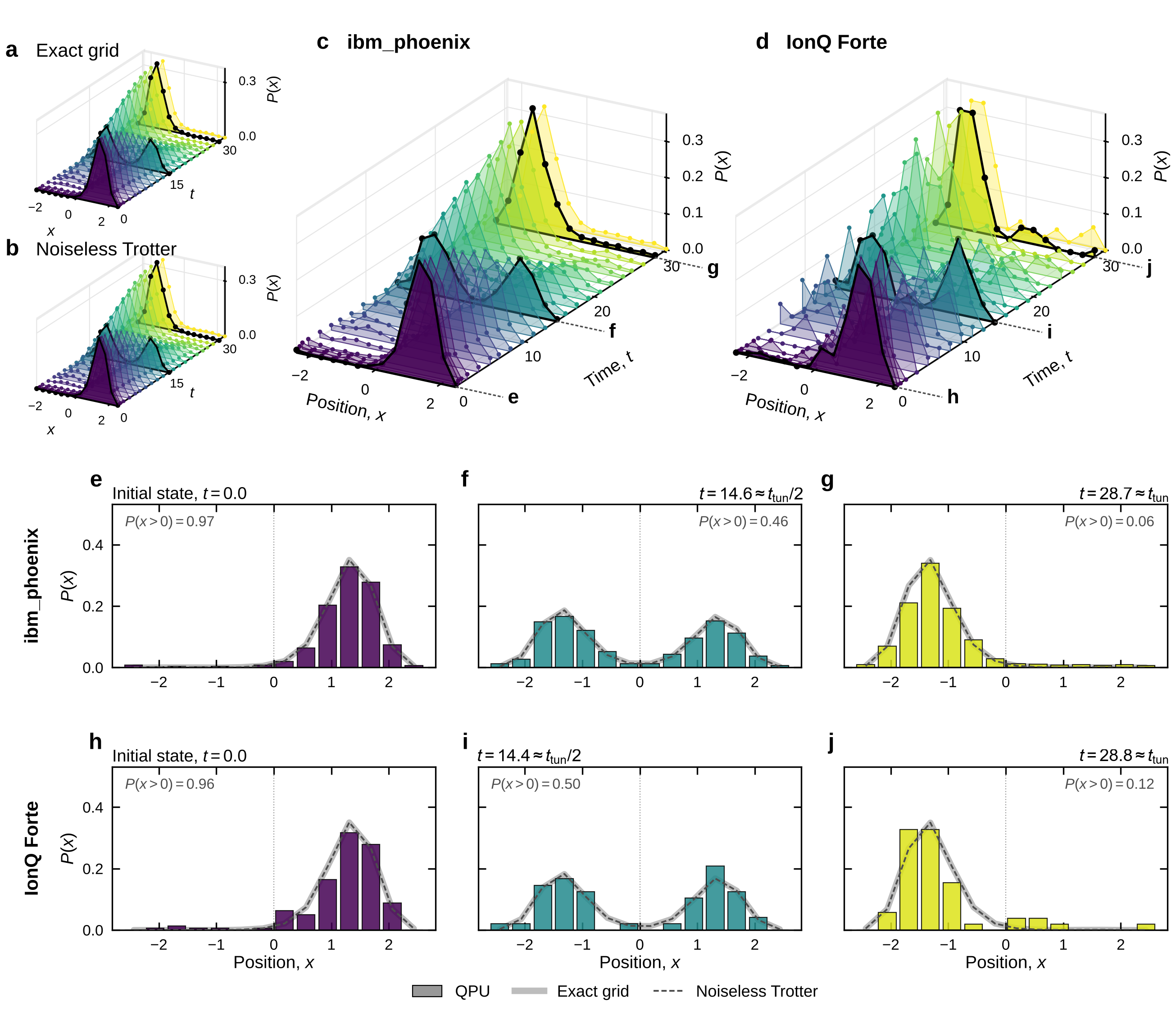}
\caption{\textbf{Double-well probability dynamics with 14-site unary encoding.}
Waterfall plots show (a) exact-grid, (b) ideal product-formula,
(c) IBM Phoenix, and (d) IonQ Forte results during the first interwell transfer.
Snapshots (e--g) and (h--j) show the corresponding IBM Phoenix and
IonQ Forte distributions at the initial, intermediate, and transfer times.
Colored bars represent measured site probabilities conditioned on
single-excitation acceptance, while gray filled and dashed profiles
denote exact-grid and ideal product-formula predictions, respectively.
Color indicates time, and $P(x>0)$ denotes the right-well probability.}
    \label{fig:dw_overview3d_first_transfer}
\end{figure}

Figure~\ref{fig:dw_overview3d_first_transfer} displays the site distributions underlying these moments. IBM snapshots are taken at $t=0$, $t\simeq14.6$, and
$t\simeq28.7$, with the nonzero acquisition times rounded
to one decimal place for display. The analysis uses the
unrounded acquisition times. The corresponding conditional
right-well probabilities are
$P_R\equiv P(x>0)=\sum_{x_j>0}\hat P_j=0.973$, $0.461$ and $0.061$, with accepted pooled weights 2,011, 1,165 and 1,162, respectively. IonQ snapshots occur at $0$, $14.4$ and $28.8$, with probabilities 0.962, 0.500 and 0.115 and accepted provider-derived weights 158, 48 and 52. No interpolation of IBM data to IonQ times is made. The redistribution from a right-well peak through two-well support to a predominantly left-well peak supports conditional population transfer on both devices. The displayed probabilities are per site, and residual distributional differences remain. Position measurements alone do not determine the accepted density matrix.

The IBM binary and unary data are assessed against their respective
finite-grid references. Their common potential does not make the two
discrete Hamiltonians identical. Likewise, the IBM and IonQ unary
observations must be interpreted together with their shot budgets
and retained weights. These comparisons characterize the complete
implementations rather than an encoding-only or processor-only effect.

The well populations also constrain coherence between energy
eigenstates. Under the ideal position-measurement model, let
$\rho$ denote the normalized state conditioned on
single-excitation post-selection.
Let $\{\ket{\phi_n}\}$ be the eigenstates of the fourteen-site
unary Hamiltonian, and write
$\rho=\sum_{m,n}\rho_{mn}\ket{\phi_m}\bra{\phi_n}$.
The right-well projector on $\mathcal H_1$ is
$\hat\Pi_R=\sum_{x_j>0}\hat n_j$, so that
$P_R=\operatorname{Tr}(\rho\hat\Pi_R)$.
Separating the diagonal and off-diagonal elements gives
\begin{equation}
    P_R=\sum_n\rho_{nn}\,
    \underbrace{\bra{\phi_n}\hat\Pi_R\ket{\phi_n}}_{=\,1/2}
    \;+\;\sum_{m\neq n}\rho_{mn}\bra{\phi_n}\hat\Pi_R\ket{\phi_m}.
    \label{eq:PR_decomposition}
\end{equation}
The grid and the potential are reflection symmetric, the grid has no
site at $x=0$, and the spectrum of the tridiagonal Hamiltonian is
nondegenerate. Each eigenstate therefore has definite parity, and
$|\phi_n(x_j)|^2$ is symmetric under $x_j\to-x_j$, so that every
eigenstate populates the two wells equally. With
$\sum_n\rho_{nn}=1$, Eq.~\eqref{eq:PR_decomposition} becomes
\begin{equation}
    P_R-\frac12=\sum_{m\neq n}\rho_{mn}\bra{\phi_n}\hat\Pi_R\ket{\phi_m}.
    \label{eq:PR_offdiagonal}
\end{equation}
The populations $\rho_{nn}$ contribute exactly $1/2$, however they are
distributed among the levels, and any deviation of $P_R$ from $1/2$
comes from the coherences $\rho_{mn}$ between different energy
eigenstates. For two eigenstates of the same parity, orthogonality
implies $\bra{\phi_n}\hat\Pi_R\ket{\phi_m}=0$, so only coherences
between even and odd eigenstates contribute. In particular, a state
dephased in the energy basis has $P_R=1/2$ at all times and is not
localized in either well.

To make this quantitative, we use the Cauchy--Schwarz inequality,
\begin{equation}
    \bigl|\bra{\phi_n}\hat\Pi_R\ket{\phi_m}\bigr|
    \le\Bigl(\bra{\phi_n}\hat\Pi_R\ket{\phi_n}
    \bra{\phi_m}\hat\Pi_R\ket{\phi_m}\Bigr)^{1/2}=\frac12 .
\end{equation}
Applying it to Eq.~\eqref{eq:PR_offdiagonal} gives
\begin{equation}
    C_{l_1}(\rho)\equiv\sum_{m\neq n}|\rho_{mn}|
    \;\ge\;2\,\Bigl|P_R-\frac12\Bigr|,
    \label{eq:coherence_witness}
\end{equation}
where $C_{l_1}$ is the $l_1$ norm of coherence in the energy basis~\cite{Baumgratz:2013ecx}.

We evaluate the witness expression $B(P_R)=2|P_R-\tfrac12|$ using the
measured conditional probabilities. The point estimates are
$\widehat B=0.95$ at $t=0$ and $0.88$ at $t=28.7$ on IBM Phoenix, and
$0.92$ at $t=0$ and $0.77$ at $t=28.8$ on IonQ Forte. We obtain
one-sided 95\% lower confidence limits from 10\,000 bootstrap resamples
drawn as for the error bars, with whole joint-shot records resampled for
IBM Phoenix. They give $C_{l_1}\ge0.93$ and $0.85$ on IBM Phoenix and
$C_{l_1}\ge0.87$ and $0.62$ on IonQ Forte at the same times, so the
coherence remains nonzero at this confidence level within the ideal
position-measurement model.

For comparison, the ideal equal-weight superposition of the
lowest doublet, Eq.~\eqref{eq:dw_initial}, has $C_{l_1}=1$.
The witness concerns coherence in the full energy eigenbasis
of the fourteen-site Hamiltonian. It does not establish
confinement to the lowest doublet or quantify fidelity to
the ideal doublet superposition. Post-selection retains
measured one-hot outcomes; it does not measure individual
energy-level populations.

\subsection{Coupled-oscillator benchmarks}\label{app:coupled_oscillators}\label{sec:coupled}

We extend the position-grid benchmarks to two coupled degrees of freedom,
using the Hamiltonians
\begin{equation}
    H_W=\frac{p_x^2+p_y^2}{2}
        +\frac{k}{2}(x^2+y^2)+g_W W(x,y),
    \qquad W(x,y)=xy\ \text{or}\ x^2y^2.
    \label{eq:coupled_benchmark_hamiltonian}
\end{equation}
We use $k=0.36$, with four sites per coordinate, $\Delta x=\Delta y=1.15$, and a product-formula step size of $\Delta t=0.8$. Binary encoding uses two qubits per coordinate with periodic spectral kinetics ($L=2.3$), whereas unary encoding uses four qubits per coordinate with Dirichlet boundaries ($L=2.875$). Thus, the initial site coordinates and amplitudes are identical, while the kinetic operators and qubit requirements differ. The figures compare coordinate means and variances with the encoding-specific exact-grid and ideal product-formula predictions. The unary circuits for these benchmarks are built from the product formula and optimized classically with Algorithm~\ref{alg:unary_compression_simple} before execution. The inter-register couplings are products of $Z$ operators acting on both registers and do not commute with the exchange gates within each register, so the local relations of Sec.~\ref{sec:unary_compression} act only between successive coupling layers. Unlike the single-register circuits, the optimized coupled circuits therefore still grow with the number of Trotter steps (Table~\ref{tab:unary_gate_costs}). In the fermionic language of Sec.~\ref{sec:unary_compression}, these couplings are density--density interactions between the two registers, so the single-register fixed-size optimization does not apply to the full coupled evolution.

Each encoding and processor has six measurement times: $t=0,\,0.8,\,2.4,\,4.8,\,6.4,\,7.2$ for the $xy$ interaction and $t=0,\,0.8,\,3.2,\,4.0,\,5.6,\,6.4$ for the $x^2y^2$ interaction. Within each model, the same times are used across encodings and processors. Unary statistics are conditioned on one excitation per register, using the accepted joint-weight normalization of Eq.~\eqref{eq:review_measurement}.

The initial state is a product of two Gaussian wave packets.
We sample their amplitudes on the position grid and normalize
the resulting state,
\begin{equation}
    \psi_{ij}(0)=\mathcal N
    \exp\left[
        -\frac{(x_i-x_0)^2}{2s_x^2}
        -\frac{(y_j-y_0)^2}{2s_y^2}
    \right],
    \qquad
    \sum_{i,j=1}^{N}|\psi_{ij}(0)|^2=1,
\end{equation}
where $\mathcal N$ is the normalization factor.
The amplitudes are real and positive, and the two coordinates
are initially uncorrelated. We use the same normalized
amplitudes for both encodings and both interaction models.
In the continuum, the Gaussian means are $x_0$ and $y_0$,
with position variances $s_x^2/2$ and $s_y^2/2$.
Sampling on the finite grid changes these moments.
Table~\ref{tab:coupled_initial_parameters} therefore lists
both the Gaussian parameters and the initial moments evaluated
from the normalized grid state.

\begin{table*}[h]
\centering
\caption{Parameters and initial-state moments of the
coupled-oscillator benchmarks in units $\hbar=m=1$.
The bilinear and quartic interactions are $g_{xy}xy$ and
$g_{x^2y^2}x^2y^2$, respectively.
Here, $N$ and $Q$ denote the numbers of sites and qubits
per coordinate. Both encodings and both interaction models
use the same initial grid amplitudes.}
\label{tab:coupled_initial_parameters}
\small
\renewcommand{\arraystretch}{1.15}
\begin{tabular*}{\textwidth}
{@{\extracolsep{\fill}}lcc@{}}
\toprule
Quantity & Binary & Unary \\
\midrule
$k$ & \multicolumn{2}{c}{0.360} \\
$g_{xy}$ & \multicolumn{2}{c}{0.120} \\
$g_{x^2y^2}$ & \multicolumn{2}{c}{0.040} \\
$N$ & 4 & 4 \\
$Q$ & 2 & 4 \\
$\Delta x=\Delta y$ & \multicolumn{2}{c}{1.150} \\
$\Delta t$ & \multicolumn{2}{c}{0.800} \\
\midrule
$x_0$ & \multicolumn{2}{c}{0.250} \\
$y_0$ & \multicolumn{2}{c}{$-0.200$} \\
$s_x^2$ & \multicolumn{2}{c}{1.400} \\
$s_y^2$ & \multicolumn{2}{c}{1.600} \\
\midrule
$\langle x\rangle(0)$
    & \multicolumn{2}{c}{0.241} \\
$\langle y\rangle(0)$
    & \multicolumn{2}{c}{$-0.188$} \\
$\mathrm{Var}(x)(0)$
    & \multicolumn{2}{c}{0.670} \\
$\mathrm{Var}(y)(0)$
    & \multicolumn{2}{c}{0.749} \\
\bottomrule
\end{tabular*}
\end{table*}

\subsection{Bilinear interaction}

\begin{figure*}[htbp]
    \centering
    \includegraphics[width=\textwidth]{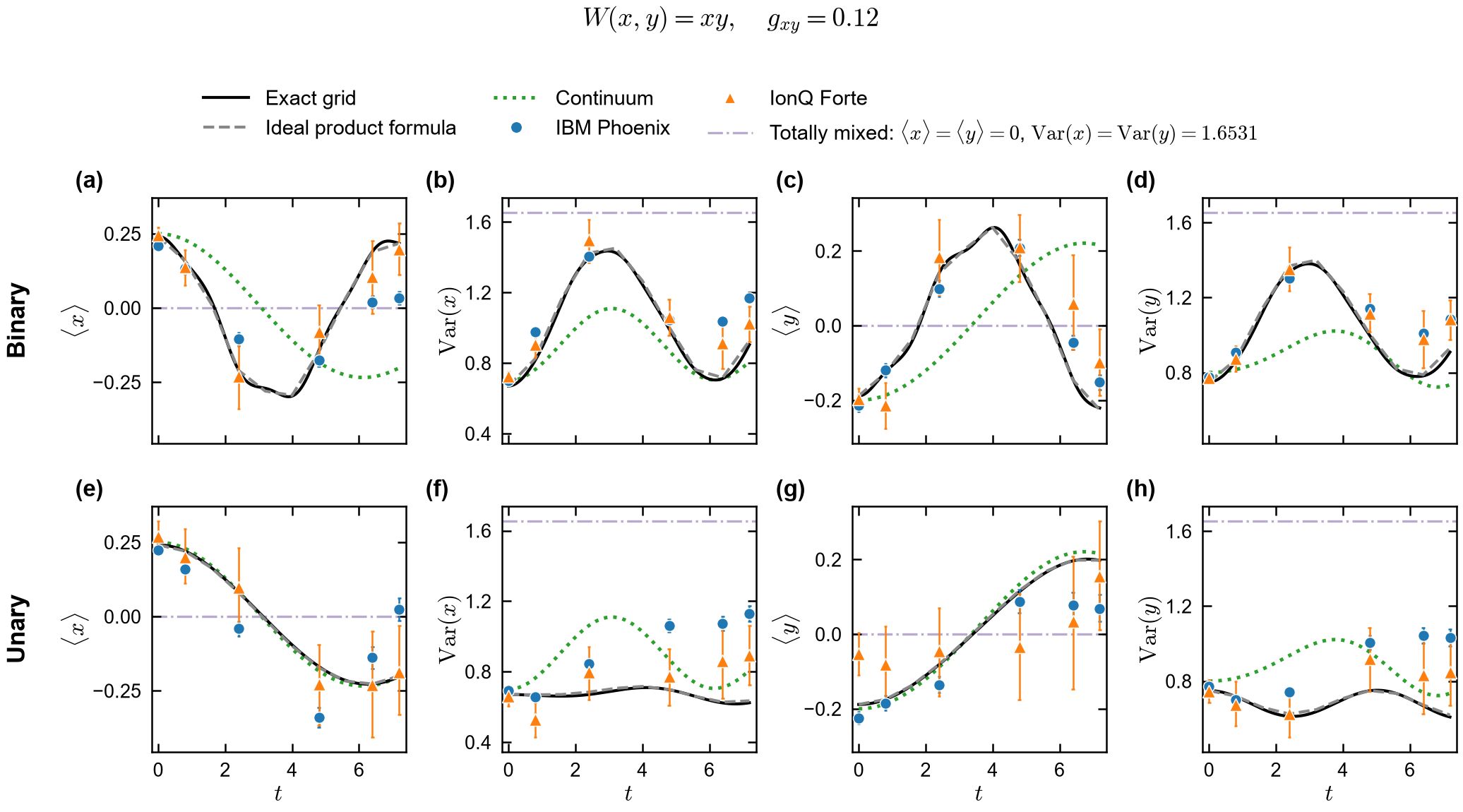}
\caption{\textbf{Dynamics of bilinearly coupled harmonic oscillators.}
Results for $k=0.36$ and $g_{xy}=0.12$, with four grid points per coordinate
and a Trotter step size of $\Delta t=0.8$.
Binary (a--d) and unary (e--h) panels show
$\braket{x}$, $\mathrm{Var}(x)$, $\braket{y}$, and $\mathrm{Var}(y)$
from left to right.
Blue circles and orange triangles denote IBM Phoenix and IonQ Forte
measurements at six sampled times up to $t=7.2$.
Black solid and gray dashed curves represent the encoding-specific
exact-grid and ideal product-formula predictions, respectively,
while green dotted curves show the exact continuum Gaussian dynamics.
Purple dash-dotted lines indicate the maximally mixed reference values,
$\braket{x}=\braket{y}=0$ and
$\mathrm{Var}(x)=\mathrm{Var}(y)=1.6531$.}
    \label{fig:coupled_xy}
\end{figure*}

For $W(x,y)=xy$, we use $g_{xy}=0.12$.
Figure~\ref{fig:coupled_xy} presents the two coordinate
means and variances for both encodings and processors.

For the bilinear model, the coarse binary exact-grid trajectories differ appreciably from the continuum Gaussian reference, so marker-to-continuum differences cannot all be attributed to hardware. The unary means can follow the continuum curves more closely, while their variances still differ. Additional QPU width growth relative to the unary exact-grid and product-formula references remains visible. Mean agreement probes the displacement of a marginal distribution and does not establish full-distribution or continuum accuracy. The same plots therefore test the physical dynamics and the chosen implementation.

\subsection{Quartic interaction}

\begin{figure*}[htbp]
    \centering
    \includegraphics[width=\textwidth]{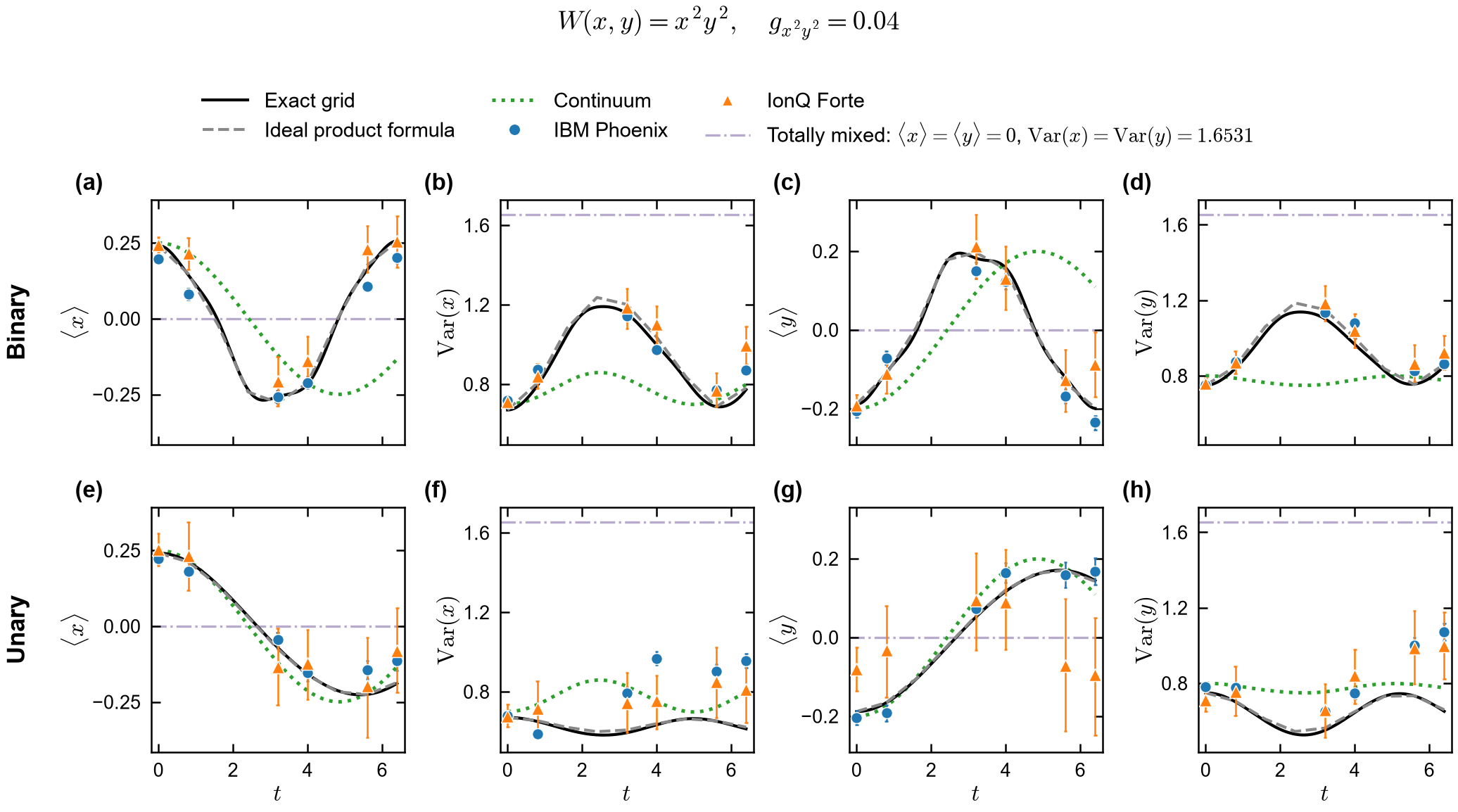}
\caption{\textbf{Dynamics of quartically coupled harmonic oscillators.}
Results for $k=0.36$ and $g_{x^2y^2}=0.04$, with four grid points per coordinate
and a Trotter step size of $\Delta t=0.8$.
Binary (a--d) and unary (e--h) panels show
$\braket{x}$, $\mathrm{Var}(x)$, $\braket{y}$, and $\mathrm{Var}(y)$
from left to right.
Blue circles and orange triangles denote IBM Phoenix and IonQ Forte
measurements at six sampled times up to $t=6.4$.
Black solid and gray dashed curves represent the encoding-specific
exact-grid and ideal product-formula predictions, respectively,
while green dotted curves show the continuum reference predictions.
Purple dash-dotted lines indicate the maximally mixed reference values,
$\braket{x}=\braket{y}=0$ and
$\mathrm{Var}(x)=\mathrm{Var}(y)=1.6531$.}
    \label{fig:coupled_x2y2}
\end{figure*}

For $W(x,y)=x^2y^2$, we use $g_{x^2y^2}=0.04$.
Figure~\ref{fig:coupled_x2y2} uses the same panel layout
and reference conventions for this interaction. The numerical
continuum reference uses Fourier Strang evolution on a $192^2$-point grid over $[-10,10)^2$ with an internal time step $\delta t=0.004$.
At $t=0,3.2,6.4$, the four coordinate moments differ by less than
$9\times10^{-7}$ from a refinement using a $256^2$-point grid over
$[-12,12)^2$ with $\delta t=0.002$.

For the quartic interaction, separation of exact-grid and ideal-product-formula variances at $\Delta t=0.8$, particularly for binary encoding, exposes time-step error before hardware is included. The displacement of a QPU marker from the product-formula curve is a separate implementation discrepancy. Unary means can track their references while late-time variances increase, again showing that a small mean error need not imply a well-preserved distribution. Neither encoding is uniformly closer across all observables and references. The present data do not isolate a specific hardware noise mechanism.

\FloatBarrier

\subsection{Gate costs}

\begin{table*}[!htbp]
\centering
\caption{Two-qubit circuit resources for selected QPU experiments.
IBM entries describe the selected region $r_0$, rather than the complete
packed circuit; routing costs may differ among regions. Counts include
state preparation and time evolution, while depth denotes
entangling-gate layers.
SHO uses $N=8$, while DW uses $N=16$ (binary)
and $N=14$ (unary).
Coupled oscillators use four grid sites per coordinate,
with interactions proportional to $xy$ or $x^2y^2$.}
\label{tab:unary_gate_costs}
\footnotesize
\setlength{\tabcolsep}{0pt}
\renewcommand{\arraystretch}{1.1}
\begin{tabular*}{\textwidth}
{@{\extracolsep{\fill}}lllccrr@{}}
\toprule
System & Processor & Encoding & $t\,(\Delta t)$
& 2Q gate & 2Q count & 2Q depth \\
\midrule
SHO & IBM Phoenix & Binary & $16\,(0.4)$
    & CZ/$R_{ZZ}$ & $934$ & $934$ \\
SHO & IBM Phoenix & Unary & $16\,(0.1)$
    & CZ & $70$ & $40$ \\
SHO & IonQ Forte & Binary & $13.6\,(0.4)$
    & ZZ & $347$ & $347$ \\
SHO & IonQ Forte & Unary & $16\,(0.1)$
    & ZZ & $70$ & $40$ \\
\midrule
DW & IBM Phoenix & Binary & $26.4\,(0.2)$
   & CZ/$R_{ZZ}$ & $7\,186$ & $7\,186$ \\
DW & IBM Phoenix & Unary & $30.4\,(0.0138)$
   & CZ & $208$ & $76$ \\
DW & IonQ Forte & Binary & $26.4\,(0.2)$
   & ZZ & $3\,320$ & $2\,393$ \\
DW & IonQ Forte & Unary & $30.4\,(0.0138)$
   & ZZ & $182$ & $50$ \\
\midrule
Coupled HO ($xy$)
   & IBM Phoenix & Binary & $7.2\,(0.8)$
   & CZ/$R_{ZZ}$ & $155$ & $155$ \\
Coupled HO ($xy$)
   & IBM Phoenix & Unary & $7.2\,(0.8)$
   & CZ/$R_{ZZ}$ & $745$ & $442$ \\
Coupled HO ($xy$)
   & IonQ Forte & Binary & $7.2\,(0.8)$
   & ZZ & $80$ & $51$ \\
Coupled HO ($xy$)
   & IonQ Forte & Unary & $7.2\,(0.8)$
   & ZZ & $352$ & $146$ \\
\midrule
Coupled HO ($x^2y^2$)
   & IBM Phoenix & Binary & $6.4\,(0.8)$
   & CZ/$R_{ZZ}$ & $67$ & $67$ \\
Coupled HO ($x^2y^2$)
   & IBM Phoenix & Unary & $6.4\,(0.8)$
   & CZ/$R_{ZZ}$ & $663$ & $395$ \\
Coupled HO ($x^2y^2$)
   & IonQ Forte & Binary & $6.4\,(0.8)$
   & ZZ & $31$ & $20$ \\
Coupled HO ($x^2y^2$)
   & IonQ Forte & Unary & $6.4\,(0.8)$
   & ZZ & $316$ & $131$ \\
\bottomrule
\end{tabular*}
\par\smallskip
\begin{minipage}{\textwidth}
\footnotesize\raggedright
For IBM circuits, the native two-qubit gates used are listed
explicitly. When both CZ and $R_{ZZ}$ gates are present,
the 2Q count is their sum, and the 2Q depth includes both types.
Resources are reported for the circuit at the last
time point shown for each data set.
\end{minipage}
\end{table*}

Table~\ref{tab:unary_gate_costs} lists the two-qubit resources of the circuits at the last time point of each data set. The binary counts grow linearly with the number of Trotter steps. For a single degree of freedom, the unary counts are bounded independently of the evolution time (Sec.~\ref{sec:unary_compression}). For the harmonic oscillator, the unary circuits use 70 two-qubit gates, against 934 (IBM) and 347 (IonQ) for the binary circuits. For the double well, they use 35 (IBM) and 18 (IonQ) times fewer two-qubit gates than the binary circuits, with two-qubit depths of 76 and 50 against $7\,186$ and $2\,393$.

For the coupled oscillators the comparison is reversed. The inter-register couplings prevent applying the same fixed-size optimization across the complete Trotterized evolution, and on the four-site grids the binary registers need only two qubits per coordinate. The binary circuits are therefore shorter by a factor of about five for the bilinear coupling and about ten for the quartic coupling. The resource trade-off depends in part on whether the fixed-size single-register optimization applies to the complete evolution.

The double-well entries of Table~\ref{tab:unary_gate_costs} compare
binary product-formula circuits with optimized unary circuits.
Appendix~\ref{app:compilation} additionally reports noiseless
product-formula baselines for both encodings, with the unary
circuits evaluated without the fixed-size optimization. For unary encoding at
$\Delta t=0.1$ and $t=28.1$, the logical construction uses
$7\,346$ two-qubit gates with a two-qubit depth of $1\,152$.
For binary encoding at $\Delta t=0.2$ and $t=26.6$, the offline
native-basis compilation gives $7\,141$ gates with depth $7\,141$
for IBM Phoenix ($R_{ZZ}$ and CZ), and $3\,345$ gates with depth
$2\,411$ for the IonQ ZZ basis.
These resource estimates include preparation, but use different
finite-grid Hamiltonians and the counting conventions specified
in the appendix. The hardware entries of
Table~\ref{tab:unary_gate_costs} refer to their original acquisition
times and compilation pipelines.

\section{Discussion}\label{sec:discussion}

The benchmarks illustrate the resource and representation choices involved in position-grid simulation. Binary encoding stores $N$ sites in $\log_2N$ qubits and uses Fourier transforms for the chosen spectral kinetic operator. Unary encoding uses one qubit per site, with neighboring-site hopping and local $Z$ rotations for the potential of a single degree of freedom. The spectral kinetic operator is exact for its finite-grid definition; neither implementation is an exact representation of the continuum Hamiltonian. Comparisons must therefore account for the kinetic discretization and boundary conditions as well as the qubit encoding.

For a single degree of freedom, the unary dynamics is a free-fermion problem, and its product formula optimizes exactly into a circuit of fixed size. With this compilation the measured double-well dynamics follows the exact-grid transfer on both processors, whereas the binary product formula, whose size grows with the number of steps, does not reproduce the transfer. The noiseless baselines in Appendix~\ref{app:compilation} quantify the trade-off between product-formula accuracy and circuit resources for the chosen discretizations. They do not by themselves establish that compilation or encoding alone caused the observed hardware contrast. The experimental comparisons concern complete implementations, including their finite-grid Hamiltonians, compilation procedures and post-selection.

Single-excitation post-selection retains measured one-hot outcomes but does not detect errors within the encoding space. Within the ideal position-measurement model, Eq.~\eqref{eq:coherence_witness} nevertheless bounds the coherence between energy eigenstates in the conditional state: after the first transfer, $C_{l_1}\ge0.85$ on IBM Phoenix and $C_{l_1}\ge0.62$ on IonQ Forte at 95\% confidence. The bound does not establish confinement to the lowest doublet or fidelity to the ideal doublet superposition.

The same free-fermion reduction to a fixed-size circuit does not apply to the complete coupled evolution, because the inter-register terms are interactions between fermionic modes. On their four-site grids the binary circuits are shorter and neither encoding is uniformly closer to its references. The choice of encoding therefore depends on the grid size, the evolution time and the connectivity of the device.

For several degrees of freedom, the coupled-oscillator benchmarks in Sec.~\ref{app:coupled_oscillators} show how register interactions enter the implementation. A unary representation with $N$ sites for each of $K$ degrees of freedom requires $KN$ qubits. Interactions diagonal in position become products of occupation operators across the participating registers. In particular, $x_a^2x_b^2$ couples two registers; four registers are involved only for terms containing four distinct degrees of freedom. Their gate costs depend on both the interaction structure and connectivity. Extending these benchmarks requires control of the field range, grid spacing and time step alongside native circuit resources and the measured weight retained after post-selection.

\backmatter

\bmhead{Acknowledgements}

We thank Junghee Ryu and Enrico Rinaldi for valuable discussions. J.Y. was supported by the National Research Foundation of Korea (NRF) grant funded by the Korean government (MSIT) (RS-2022-NR069038). J.Y. and S.S. were supported by the National Research Foundation of Korea (NRF) grant funded by the Korean government (MSIT) (RS-2025-25466315). J.Y. was supported by Brain Pool program funded by the Ministry of Science and ICT through the National Research Foundation of Korea (RS-2025-25457100). This work was partly supported by the Institute of Information \& Communications Technology Planning \& Evaluation(IITP)-ITRC(Information Technology Research Center) grant funded by the Korea government(MSIT)(IITP-2026-RS-2026-25519864). This work was supported by Creation of the Quantum Information Science R\&D Ecosystem (Grant No. RS-2023-NR068116) through the National Research Foundation of Korea (NRF) funded by the  Korean government (Ministry of Science and ICT). We acknowledge the Yonsei University Quantum Computing Project Group for providing support and access to the Quantum System One, which is operated at Yonsei University. This research was supported by ‘Creation of the quantum information science R\&D ecosystem(based on human resources)’ through the National Research Foundation of Korea(NRF) funded by the Korean government (RS-2023-NR057243). This work was supported by quantum computing cloud resources provided by the Korea Institute of Science and Technology Information (KISTI) through the National Research Foundation of Korea (NRF) grant funded by the Korea government (MSIT) (No. RS-2025-24534879).




\begin{appendices}
\renewcommand{\theHequation}{\thesection.\arabic{equation}}
\counterwithin*{table}{section}
\counterwithin*{figure}{section}

\section{Gaussian wave packet in the harmonic oscillator}\label{app:gauss}

For the initial Gaussian in Eq.~\eqref{eq:ho_initial}, the
oscillator propagator follows from
$K(x,t;x',0)=\sum_{n=0}^{\infty}\phi_n(x)\phi_n^*(x')
e^{-\mathrm{i}\omega(n+1/2)t}$. Summing the Hermite-polynomial series
gives Mehler's kernel. For $0<t<\pi/\omega$,
\begin{equation}
    K(x,t;x',0)=\sqrt{\frac{\omega}{2\pi \mathrm{i}\sin(\omega t)}}
    \exp\!\left\{\frac{\mathrm{i}\omega}{2\sin(\omega t)}
    \left[(x^2+x'^2)\cos(\omega t)-2xx'\right]\right\}.
    \label{eq:app_mehler}
\end{equation}
At later times the kernel is defined by unitary continuation
through the points $t=n\pi/\omega$. Integrating it against
Eq.~\eqref{eq:ho_initial} yields
\begin{equation}
\begin{aligned}
    \psi(x,t)
    &=(\pi s^2)^{-1/4}
      \left[\cos(\omega t)+\frac{\mathrm{i}\sin(\omega t)}{\omega s^2}
      \right]^{-1/2}\\
    &\quad\times\exp\!\left\{
    \frac{-x^2[\cos(\omega t)+\mathrm{i}\omega s^2\sin(\omega t)]
          +2x_0x-x_0^2\cos(\omega t)}
         {2[s^2\cos(\omega t)+\mathrm{i}\sin(\omega t)/\omega]}
    \right\}.
\end{aligned}
\label{eq:app_gaussian_evolution}
\end{equation}
The square-root factor is continued continuously from its
value $1$ at $t=0$; choosing its principal branch independently
at each time would introduce phase jumps. The initial limit
recovers Eq.~\eqref{eq:ho_initial}. For $s^2=1/\omega$,
\begin{equation}
    |\psi(x,t)|^2=\sqrt{\frac{\omega}{\pi}}
        \exp[-\omega(x-x_0\cos\omega t)^2],
    \label{eq:app_coherent_density}
\end{equation}
so the packet translates without changing its width.

The moments can also be obtained from the Heisenberg equations.
They give $\mathrm d^2\langle\hat x\rangle/\mathrm dt^2
=-\omega^2\langle\hat x\rangle$ and
\begin{equation}
    \frac{\mathrm d^2}{\mathrm dt^2}\langle\hat x^2\rangle
        +4\omega^2\langle\hat x^2\rangle=4E,
    \qquad
    E=\frac{1}{4s^2}+\frac{\omega^2s^2}{4}
         +\frac{\omega^2x_0^2}{2}.
\end{equation}
With zero initial mean momentum and zero initial symmetrized
position--momentum covariance,
\begin{equation}
    \langle\hat x^2\rangle(t)
    =\left(x_0^2+\frac{s^2}{2}\right)\cos^2(\omega t)
     +\frac{\sin^2(\omega t)}{2\omega^2s^2}.
\end{equation}
Subtracting $\langle\hat x\rangle^2$ reproduces the variance
given with Eq.~\eqref{eq:ho-moments}.

\section{Discretization error}\label{sec:discerr}
\begin{figure*}[!tp]
    \centering
    \includegraphics[width=\textwidth]{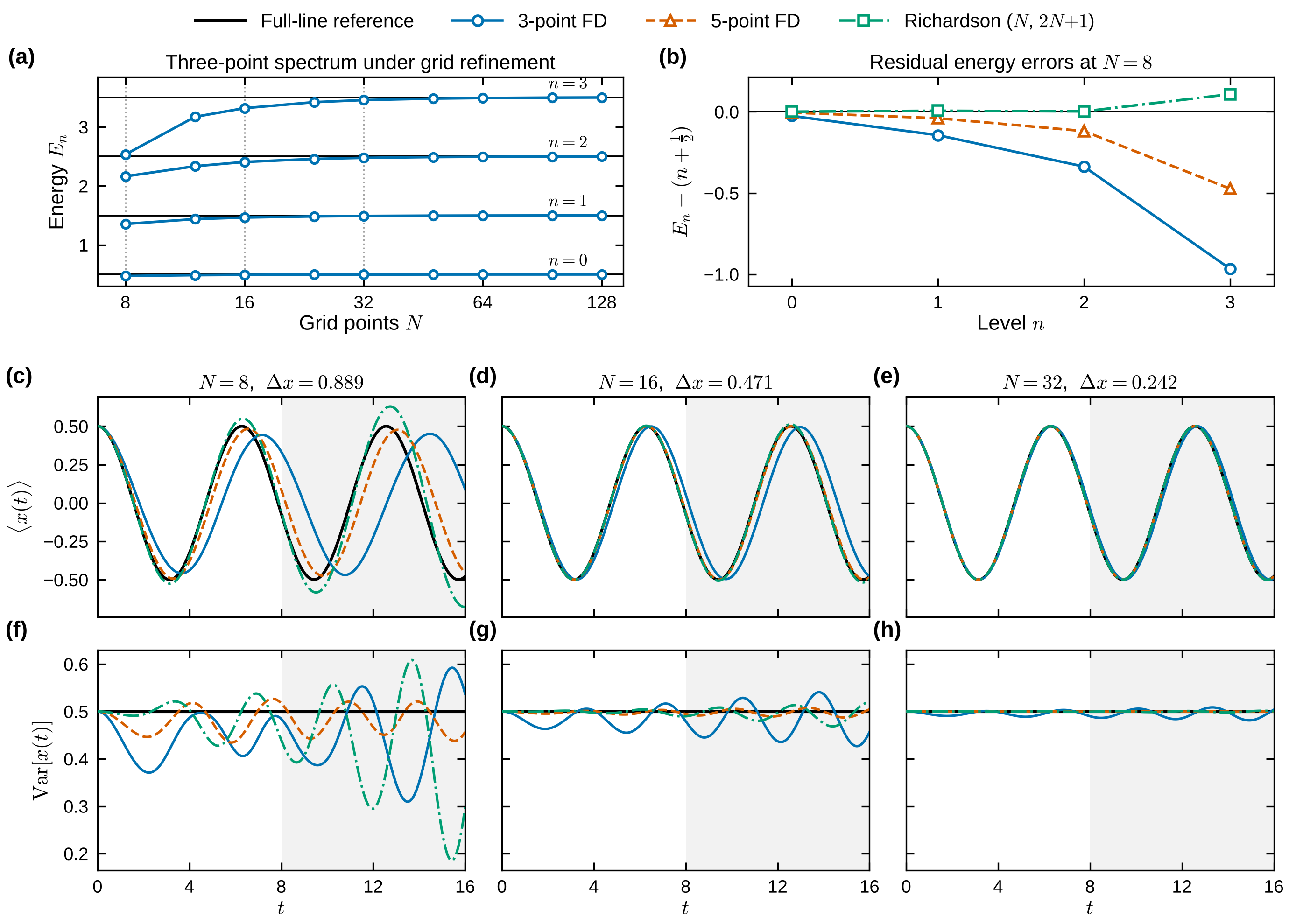}
\caption{\textbf{Grid-size dependence of the harmonic-oscillator spectrum and dynamics.}
The interval $x\in[-4,4]$ is discretized into $N$ interior grid points
with Dirichlet boundary conditions and spacing $\Delta x=8/(N+1)$,
using $\hbar=m=\omega=1$.
(a) Lowest four energy levels of the three-point finite-difference
Hamiltonian as a function of $N$, with horizontal lines indicating
the full-line energies $E_n=n+\tfrac12$.
(b) Energy deviations $E_n-(n+\tfrac12)$ at $N=8$, comparing
three-point and five-point finite differences with Richardson extrapolation.
(c--e) Mean position and (f--h) position variance for $N=8,16,32$,
respectively, initialized from a Gaussian centered at $x_0=0.5$.
Black curves indicate the full-line references
$\langle x(t)\rangle=0.5\cos t$ and $\mathrm{Var}(x)=1/2$.
The five-point stencil improves the coarse-grid dynamics relative to
the three-point stencil, while increasing $N$ further reduces the deviations.
Gray shading marks the late-time interval $8\le t\le16$.}

\label{fig:sho_discretization_v3}
\end{figure*}

Figure~\ref{fig:sho_discretization_v3} illustrates ideal spatial-discretization effects for $H=(p^2+x^2)/2$ in units $\hbar=m=\omega=1$, using $N=Q$ retained Dirichlet sites on $(-4,4)$ and spacing $\Delta x=8/(N+1)$. The initial Gaussian has center $x_0=0.5$ and $s^2=1$; it is sampled and renormalized on each grid. These illustrative parameters differ from the experimental SHO settings in Table~\ref{tab:sho_initial_parameters}. Its finite-Hamiltonian evolution contains neither product-formula error nor QPU noise, and its full-line reference is the coherent-state solution $\langle x\rangle=0.5\cos t$, $\operatorname{Var}(x)=1/2$.

The three-point finite difference used for unary evolution
and the Fourier kinetic operator used for binary evolution
are different numerical approximations. The distinction is
not an intrinsic restriction of either encoding. The Fourier
operator is exact as a finite-grid matrix; approximation of
continuum dynamics additionally requires control of the
field range, periodic boundary effects and unresolved
momenta or aliasing.

For smooth states with negligible boundary effects, the
three-point finite difference in
Eq.~\eqref{eq:unary_laplacian} gives
\begin{equation}
    \hat p^2_{\mathrm{FD}}
    = \hat p^2-\frac{\Delta x^2}{12}\hat p^4
      +O(\Delta x^4).
\end{equation}
The leading correction to the kinetic energy is therefore
$-\Delta x^2\hat p^4/24$.
For the harmonic oscillator, first-order perturbation theory
then gives
\begin{align}
    E_n(\Delta x)
       &=\omega\left(n+\frac12\right)
         -\frac{\omega^2\Delta x^2}{32}(2n^2+2n+1)
         +O(\Delta x^4),\notag\\
    E_n-E_{n-1}
       &=\omega-\frac{n\omega^2\Delta x^2}{8}
         +O(\Delta x^4),\qquad n\ge1.
    \label{eq:lattice-shift}
\end{align}
These expressions assume that finite-box effects on the
states considered are negligible. 

Equation~\eqref{eq:lattice-shift} shows that the leading
finite-difference correction to the oscillator energies
is of order $\Delta x^2$. Reducing the grid spacing
suppresses this error, while the field range $L$ must
be large enough to suppress boundary effects over the
evolution interval considered. These two sources of
error must therefore be controlled separately.
For the unary Dirichlet grid,
$Q=2L/\Delta x-1$, so improving the spatial resolution
at fixed field range requires more qubits.

\subsection{Improved kinetic operators}\label{sec:improved}

Combining three-point formulas at spacings $\Delta x$ and
$2\Delta x$ cancels the leading bulk error and gives
\begin{equation}
    f''(x)=\frac{-f(x+2\Delta x)+16f(x+\Delta x)-30f(x)
                    +16f(x-\Delta x)-f(x-2\Delta x)}
                   {12\Delta x^2}+O(\Delta x^4).
\end{equation}
Writing $S_{j,k}=(X_jX_k+Y_jY_k)/2$, the corresponding
single-excitation stencil matrix is
\begin{equation}
\begin{aligned}
    \hat p^2_{\mathrm{5pt}}
       &=-\frac{4}{3\Delta x^2}\sum_{j=1}^{Q-1}S_{j,j+1}
         +\frac{1}{12\Delta x^2}\sum_{j=1}^{Q-2}S_{j,j+2}\\
       &\quad+\frac{5}{2\Delta x^2}\sum_{j=1}^{Q}\hat n_j.
\end{aligned}
\label{eq:p5pt}
\end{equation}
For the same zero extension outside the retained grid,
the fourth-derivative stencil is
\begin{equation}
    \hat p^4_{\mathrm{FD}}=\frac{1}{\Delta x^4}
    \left[6\sum_{j=1}^{Q}\hat n_j
        -4\sum_{j=1}^{Q-1}S_{j,j+1}
        +\sum_{j=1}^{Q-2}S_{j,j+2}\right].
    \label{eq:app_fourth_derivative}
\end{equation}
The matrix identity
$\hat p^2_{\mathrm{5pt}}=\hat p^2_{\mathrm{FD}}
+\Delta x^2\hat p^4_{\mathrm{FD}}/12$
then reproduces Eq.~\eqref{eq:p5pt}.
Dirichlet conditions specify the two boundary values, but
do not specify the additional exterior values required by
the five-point stencil. The displayed zero extension is
therefore not, by itself, a globally fourth-order boundary
closure. Fourth-order convergence of low-energy eigenvalues
additionally requires a suitable closure or boundary effects
below the bulk discretization error.

Richardson extrapolation offers an alternative using the
nearest-neighbor stencil. At the same $L$, grids with $Q$
and $2Q+1$ sites have spacings $\Delta x$ and $\Delta x/2$.
If an observable at fixed physical time has an expansion
$\mathcal O(\Delta x)=\mathcal O(0)+c_2\Delta x^2
+c_4\Delta x^4+\cdots$, then
\begin{equation}
    \mathcal O_{\mathrm R}
       =\frac{4\mathcal O(\Delta x/2)-\mathcal O(\Delta x)}{3}
       =\mathcal O(0)+O(\Delta x^4).
    \label{eq:app_richardson}
\end{equation}
The two calculations must use consistent physical states,
observables and times. This removes the leading spacing
error under the stated expansion; it does not remove
finite-box effects, product-formula error or hardware noise.

\FloatBarrier

\section{Double-well initial state}\label{app:dw_initial}

The potential in Eq.~\eqref{eq:potentials} can be written
in dimensionless form using
\begin{equation}
    u=\frac{x}{a},\qquad
    \lambda=2a^2V_0,\qquad \epsilon=2a^2E.
\end{equation}
The continuum eigenvalue equation becomes
\begin{equation}
    -\frac{\mathrm d^2\psi}{\mathrm du^2}
       +\lambda(u^2-1)^2\psi=\epsilon\psi.
    \label{eq:app_dw_dimensionless}
\end{equation}
For the parameters in Table~\ref{tab:dw_initial_parameters},
$\lambda=13.056$ to three decimal places.
The finite-grid doublets used to construct the initial
states are summarized in Table~\ref{tab:app_dw_doublet}.
Each encoding uses the two lowest eigenvectors of its own
discretized Hamiltonian in Eq.~\eqref{eq:dw_initial}.
The numbers are finite-grid predictions, not continuum
convergence or hardware results.

\begin{table}[h]
    \centering
    \caption{Finite-grid doublets for the double-well initial
    states. Energies and transfer times are evaluated from
    unrounded stored parameters and displayed to three decimals.
    The binary row describes the prepared configuration.}
    \label{tab:app_dw_doublet}
    \small
    \begin{tabular*}{\linewidth}{@{\extracolsep{\fill}}lccccc@{}}
        \toprule
        Encoding & $N$ & $E_0$ & $E_1$ & $\Delta E$ & $t_{\mathrm{tun}}$ \\
        \midrule
        Binary & 16 & 1.410 & 1.528 & 0.118 & 26.655 \\
        Unary & 14 & 1.380 & 1.492 & 0.112 & 28.088 \\
        \bottomrule
    \end{tabular*}
\end{table}

For the real even and odd eigenvectors on these symmetric
even-site grids, the ideal doublet evolution gives
\begin{equation}
    P_R(t)=\frac12+C\cos(\Delta E t),\qquad
    C=\sum_{x_j>0}\phi_0(x_j)\phi_1(x_j).
    \label{eq:app_dw_probability}
\end{equation}
The odd eigenvector is signed so that $C>0$.
The coefficient $C$ need not equal $1/2$, so localization
in a single well need not be perfect. The transfer and
return times remain $\pi/\Delta E$ and $2\pi/\Delta E$.

\section{Coefficients of powers of the binary position operator}
\label{app:binary_position}\label{bianry_pos}

For the centered grid in Eq.~\eqref{eq:binary_pos}, define
$\beta_i=(\Delta x/2)2^i$, $i=0,\ldots,Q-1$.
Using an auxiliary variable $z$, the generating function is
\begin{align}
    \mathcal G(z)=e^{z\hat x}
       &=\prod_{i=0}^{Q-1}e^{-z\beta_iZ_i}\notag\\
       &=\prod_{i=0}^{Q-1}
           [\cosh(\beta_i z)I-\sinh(\beta_i z)Z_i].
    \label{eq:position_generating_function}
\end{align}
Consequently,
\begin{equation}
    \hat x^n=\sum_{S\subseteq\{0,\ldots,Q-1\}}
        c_S^{(n)}Z_S,\qquad Z_S=\prod_{i\in S}Z_i,
    \quad Z_{\varnothing}=I,
\end{equation}
where the coefficients are
\begin{equation}
    c_S^{(n)}=
    \left.\frac{\mathrm d^n}{\mathrm dz^n}
    \left[\prod_{i\notin S}\cosh(\beta_i z)
          \prod_{i\in S}[-\sinh(\beta_i z)]\right]
    \right|_{z=0}.
    \label{eq:binary_position_coefficients}
\end{equation}
Every factor associated with $i\in S$ contributes at least
one odd power of $z$. Thus $c_S^{(n)}=0$ unless
$|S|\le\min(n,Q)$ and $|S|\equiv n\pmod 2$.
For example,
\begin{equation}
    \hat x^2=\left(\sum_i\beta_i^2\right)I
          +2\sum_{i<j}\beta_i\beta_jZ_iZ_j,
\end{equation}
recovering the quadratic potential expansion in the main text.
The same coefficient formula covers the quartic term without
requiring a separate generating function.

\section{Measurement-shot budgets}
\label{app:regional_shot_budget}

Tables~\ref{tab:ibm_region_shots}--\ref{tab:ionq_region_times} list preselection
budgets for the original hardware-result figures. Each region encodes the
complete benchmark, including both coordinates for coupled oscillators.
For $K$ acquired times with $N_r$ shots per time, region $r$ contributes $K N_r$
observations. Pooled totals sum regions and times; simultaneous regions share
the same physical shots. Region indices are zero-based and local to each
benchmark. Moments and distribution snapshots reuse the same measurements.

\begin{table}[!htbp]
\centering
\caption{IBM Phoenix budgets resolved by individual region. $K$ is the number of acquired times. The last column gives the total of each individual region over time; the pooled row then sums all regions of the same calculation. For example, the eight SHO unary regions each provide $41\,984$ observations, giving $335\,872$ pooled regional observations from $41\,984$ shared circuit shots.}
\label{tab:ibm_region_shots}\label{tab:measurement_shots}
\begingroup
\fontsize{9}{11}\selectfont
\setlength{\tabcolsep}{4pt}
\renewcommand{\arraystretch}{1.12}
\begin{tabular*}{\linewidth}{@{\extracolsep{\fill}}lllrrr@{}}
\toprule
Model & Encoding & Region & $K$ & Shots/time & Region total \\
\midrule
SHO & Binary & $r_{0}$ & 41 & 5\,200 & 213\,200 \\
 &  & $r_{1}$ & 41 & 5\,200 & 213\,200 \\
\multicolumn{5}{r}{Pooled regional observations (total)} & 426\,400 \\
\midrule
\addlinespace[3pt]
SHO & Unary & $r_{0}$ & 41 & 1\,024 & 41\,984 \\
 &  & $r_{1}$ & 41 & 1\,024 & 41\,984 \\
 &  & $r_{2}$ & 41 & 1\,024 & 41\,984 \\
 &  & $r_{3}$ & 41 & 1\,024 & 41\,984 \\
 &  & $r_{4}$ & 41 & 1\,024 & 41\,984 \\
 &  & $r_{5}$ & 41 & 1\,024 & 41\,984 \\
 &  & $r_{6}$ & 41 & 1\,024 & 41\,984 \\
 &  & $r_{7}$ & 41 & 1\,024 & 41\,984 \\
\multicolumn{5}{r}{Pooled regional observations (total)} & 335\,872 \\
\midrule
\addlinespace[3pt]
DW & Binary & $r_{0}$ & 12 & 2\,048 & 24\,576 \\
 &  & $r_{1}$ & 12 & 2\,048 & 24\,576 \\
\multicolumn{5}{r}{Pooled regional observations (total)} & 49\,152 \\
\midrule
\addlinespace[3pt]
DW & Unary & $r_{0}$ & 74 & 1\,024 & 75\,776 \\
 &  & $r_{1}$ & 74 & 1\,024 & 75\,776 \\
 &  & $r_{2}$ & 74 & 1\,024 & 75\,776 \\
\multicolumn{5}{r}{Pooled regional observations (total)} & 227\,328 \\
\midrule
\addlinespace[3pt]
Coupled $xy$ & Binary & $r_{0}$ & 6 & 5\,200 & 31\,200 \\
 &  & $r_{1}$ & 6 & 5\,200 & 31\,200 \\
\multicolumn{5}{r}{Pooled regional observations (total)} & 62\,400 \\
\midrule
\addlinespace[3pt]
Coupled $xy$ & Unary & $r_{0}$ & 6 & 5\,200 & 31\,200 \\
 &  & $r_{1}$ & 6 & 5\,200 & 31\,200 \\
\multicolumn{5}{r}{Pooled regional observations (total)} & 62\,400 \\
\midrule
\addlinespace[3pt]
Coupled $x^2y^2$ & Binary & $r_{0}$ & 6 & 5\,200 & 31\,200 \\
 &  & $r_{1}$ & 6 & 5\,200 & 31\,200 \\
\multicolumn{5}{r}{Pooled regional observations (total)} & 62\,400 \\
\midrule
\addlinespace[3pt]
Coupled $x^2y^2$ & Unary & $r_{0}$ & 6 & 5\,200 & 31\,200 \\
 &  & $r_{1}$ & 6 & 5\,200 & 31\,200 \\
\multicolumn{5}{r}{Pooled regional observations (total)} & 62\,400 \\
\bottomrule
\end{tabular*}
\endgroup
\end{table}
\FloatBarrier

\begin{table}[!htbp]
\centering
\caption{IonQ Forte budgets for the single measurement region, denoted $r_0$. Each total is accumulated over times within that calculation; there is no pooling across multiple regions. Shot counts are verified against completed-job metadata. Ranges are resolved into exact per-time budgets in Table~\ref{tab:ionq_region_times}.}
\label{tab:ionq_region_shots}
\begingroup
\fontsize{9}{11}\selectfont
\setlength{\tabcolsep}{4pt}
\renewcommand{\arraystretch}{1.12}
\begin{tabular*}{\linewidth}{@{\extracolsep{\fill}}lllrrr@{}}
\toprule
Model & Encoding & Region & $K$ & Shots/time & Region total \\
\midrule
SHO & Binary & $r_0$ & 8 & 140--2\,841 & 4\,782 \\
SHO & Unary & $r_0$ & 41 & 124 & 5\,084 \\
DW & Binary & $r_0$ & 8 & 104--1\,154 & 2\,410 \\
DW & Unary & $r_0$ & 22 & 192 & 4\,224 \\
Coupled $xy$ & Binary & $r_0$ & 6 & 227--3\,964 & 6\,672 \\
Coupled $xy$ & Unary & $r_0$ & 6 & 278--1\,001 & 2\,670 \\
Coupled $x^2y^2$ & Binary & $r_0$ & 6 & 495--3\,964 & 7\,491 \\
Coupled $x^2y^2$ & Unary & $r_0$ & 6 & 278--1\,001 & 2\,668 \\
\bottomrule
\end{tabular*}
\endgroup
\end{table}
\FloatBarrier

\begin{table}[!htbp]
\centering
\caption{Per-time shot counts for the nonuniform IonQ single-region series. The two vectors correspond entry by entry. Series with constant shot counts are specified directly in Table~\ref{tab:ionq_region_shots}.}
\label{tab:ionq_region_times}
\begingroup
\fontsize{9}{11}\selectfont
\setlength{\tabcolsep}{4pt}
\renewcommand{\arraystretch}{1.12}
\begin{tabular*}{\linewidth}{@{\extracolsep{\fill}}p{0.22\linewidth}p{0.34\linewidth}p{0.36\linewidth}@{}}
\toprule
Model and encoding & Acquisition times $t$ & Shots at each time \\
\midrule
SHO Binary & 0, 0.8, 4, 4.8, 8.8, 9.6, 12.8, 13.6 & 2841, 608, 321, 140, 272, 140, 140, 320 \\
DW Binary & 0, 0.8, 1.6, 4, 8.8, 13.6, 20, 26.4 & 1154, 143, 105, 105, 105, 347, 104, 347 \\
Coupled $xy$ Binary & 0, 0.8, 2.4, 4.8, 6.4, 7.2 & 3964, 957, 498, 513, 227, 513 \\
Coupled $xy$ Unary & 0, 0.8, 2.4, 4.8, 6.4, 7.2 & 1001, 280, 279, 384, 278, 448 \\
Coupled $x^2y^2$ Binary & 0, 0.8, 3.2, 4, 5.6, 6.4 & 3964, 1171, 696, 613, 495, 552 \\
Coupled $x^2y^2$ Unary & 0, 0.8, 3.2, 4, 5.6, 6.4 & 1001, 279, 278, 384, 278, 448 \\
\bottomrule
\end{tabular*}
\endgroup
\end{table}
\FloatBarrier

For each listed series, one acquisition point is recorded at
each acquired time; a provider job may contain several such
points. The regional totals therefore do not represent
additional independent circuit repetitions.

We did not enable IonQ debiasing because of the minimum
shot count required to use this option.

Table~\ref{tab:retention_ranges} reports the percentage of recorded outcomes
retained by unary post-selection. For one measurement region,
$f_r(t)=N_{\mathrm{acc},r}(t)/N_{\mathrm{raw}}(t)$ is the fraction of recorded
shots accepted by the unary code-space selection. For coupled oscillators,
both coordinate registers in that region must contain exactly one excitation.
The table gives the smallest and largest accepted fractions over the acquired
times and individual regions, expressed as percentages. These extrema are not
confidence intervals. IonQ fractions use provider-derived accepted weights.
For example, the IBM DW range means that approximately 26\%--76\% of the
shots in an individual region pass the selection, depending on region and time.
Pooled survival uses the sum of accepted regional weights divided by the sum
of their corresponding raw regional budgets; simultaneously measured regions
still share their physical shots.

\begin{table}[!htbp]\centering
\caption{Fraction of unary measurement outcomes retained after post-selection,
in percent. Minimum and maximum are taken over individual regions and acquired
times within each dataset. These are survival-rate extrema, not raw shot totals.}
\label{tab:retention_ranges}
\begin{tabular*}{\linewidth}{@{\extracolsep{\fill}}llrr@{}}
\toprule
Model & QPU & \shortstack{Minimum retained\\(\% of shots)} & \shortstack{Maximum retained\\(\% of shots)}\\
\midrule
SHO & IBM & 43.85 & 89.36 \\
SHO & IonQ & 56.45 & 91.94 \\
DW & IBM & 25.98 & 75.98 \\
DW & IonQ & 15.63 & 82.29 \\
Coupled $xy$ & IBM & 25.67 & 91.83 \\
Coupled $xy$ & IonQ & 33.71 & 92.81 \\
Coupled $x^2y^2$ & IBM & 32.15 & 91.17 \\
Coupled $x^2y^2$ & IonQ & 37.50 & 92.71 \\
\bottomrule\end{tabular*}\end{table}\FloatBarrier

\section{Compilation baselines for the double well}
\label{app:compilation}

The double-well circuits executed on hardware differ in compilation as
well as in encoding. Both use product formulas, while the unary circuits
also use the fixed-size optimization described in
Sec.~\ref{sec:unary_compression}. To examine this dependence, we
evaluate in noiseless emulation the unary product formula before that
optimization and the binary product formula at several time steps.
The product-formula results use fourteen-site unary and sixteen-site binary grids with $\Delta x=0.375$ and the stored potential coefficient $V_0\simeq2.9013$,
evaluated at $t=n\Delta t$ with $n$ the integer closest to
$t_{\mathrm{tun}}/\Delta t$. The hardware archive instead uses $\Delta x\simeq0.3748$,
as reported in Table~\ref{tab:benchmark_grids}. These baseline results are noiseless calculations.

\begin{table}[!htbp]
\centering
\caption{Noiseless second-order product formulas for the double well at
$t\simeq t_{\mathrm{tun}}$. $P_R$ is the right-well probability (corresponding exact-grid value approximately $0.008$), and $F$ the fidelity with the exact grid state. For
unary encoding, two-qubit counts and depths are logical, with two
entangling gates per exchange gate, phase-aware same-bond fusion across neighboring steps and the preparation
chain included. The intervening potential phases are retained; odd half-step
angles are not simply added through the potential. For binary encoding,
the two lines list IBM Phoenix ($R_{ZZ}$+CZ) and IonQ (ZZ) native resources,
respectively, at the $n$ and $t$ of each row. Preparation is included.
Values of $P_R$ and $F$ are rounded to three decimal places.}
\label{tab:compilation}
\small
\begin{tabular*}{\linewidth}{@{\extracolsep{\fill}}lcccccc@{}}
\toprule
Encoding & $\Delta t$ & $n$ & $P_R$ & $F$ & 2Q count & 2Q depth \\
\midrule
Unary  & 0.4  & 70  & 0.848 & 0.003 & $1\,860$  & 308 \\
Unary  & 0.2  & 140 & 0.456 & 0.531 & $3\,680$  & 588 \\
Unary  & 0.1  & 281 & 0.032 & 0.976 & $7\,346$  & $1\,152$ \\
Unary  & 0.05 & 562 & 0.009 & 0.999 & $14\,652$ & $2\,276$ \\
\midrule
Binary & 0.4  & 67  & 0.253 & 0.718 & \shortstack{$3\,607$ (IBM)\\$1\,695$ (IonQ)} & \shortstack{$3\,607$\\$1\,223$} \\
Binary & 0.2  & 133 & 0.015 & 0.996 & \shortstack{$7\,141$ (IBM)\\$3\,345$ (IonQ)} & \shortstack{$7\,141$\\$2\,411$} \\
Binary & 0.1  & 267 & 0.009 & 1.000 & \shortstack{$14\,382$ (IBM)\\$6\,695$ (IonQ)} & \shortstack{$14\,382$\\$4\,823$} \\
\bottomrule
\end{tabular*}
\end{table}

Among the cases tested in Table~\ref{tab:compilation}, the unary
product formula before fixed-size optimization needs a smaller time
step than the binary one to capture the transfer. The three-point kinetic operator has a large spectral range,
$\sim2/\Delta x^2\simeq14$, and the splitting of the odd- and even-bond
layers adds a further source of product-formula error. For the tested
steps $\Delta t=0.4$ and 0.2 the ideal unary circuit does not reproduce the
transfer even without noise. Among the tested step sizes, $\Delta t=0.1$ captures the transfer and requires about $7\,300$ two-qubit gates, comparable
to the binary circuit at $\Delta t=0.2$, with a two-qubit depth about
six times smaller in the shown IBM mappings. The original Table~\ref{tab:unary_gate_costs} entries $7\,186$/$3\,320$ instead use the original compilation pipeline on the archived grid at $n=132$, $t=26.4$; the present binary $\Delta t=0.2$ row uses $n=133$, $t=26.6$.

The exact $16\times16$ propagator of the four-qubit binary
finite-grid Hamiltonian can instead be synthesized directly.
Using the baseline grid with $\Delta x=0.375$, offline direct
synthesis of $e^{-\mathrm{i}\hat Ht}$ with \texttt{Qiskit}
at optimization level~3 gives 95 CZ gates with an all-to-all
structure and 156 CZ gates with a four-qubit linear-chain
structure at both tested times, $t=13.6$ and $26.4$.
For each case, we select the circuit with the lowest two-qubit
count among transpiler seeds $0$--$7$, using two-qubit depth
to break ties. These counts exclude initial-state preparation,
and the linear-chain structure does not represent the full
hardware target. Since the unary initial state is fixed,
a separate state-specific optimization can target the
classically computed ideal finite-grid state $\psi(t)$
using the Givens chain of Sec.~\ref{sec:prep}, including
local phase rotations. For $Q=14$, this construction requires
at most $2(Q-1)=26$ entangling gates before routing.
For the hardware circuits, optimization preserves the
product-formula evolution on $\mathcal H_1$ and leaves its
Trotter error unchanged. On IBM Phoenix, state preparation
is separate from the optimized evolution block. On IonQ Forte,
the number-conserving preparation rotations are optimized
jointly with the evolution, while the initial $X$ seed remains
outside the optimized block. The complete preparation-and-evolution
circuits contain 208 two-qubit gates on IBM Phoenix and 182
on IonQ Forte. Table~\ref{tab:dw_compilation_routes}
summarizes these resources, distinguishing direct propagator
synthesis, state-specific optimization, and the product-formula
circuits executed on hardware. Their costs refer to different
computational tasks and preparation scopes.

\begin{table}[!htbp]
\centering
\caption{Two-qubit resources for double-well circuit constructions.
Binary direct synthesis implements the exact finite-grid propagator
at $t=13.6$ and $26.4$, excluding initial-state preparation.
State-specific unary optimization targets the prescribed states
at $t=0$, $14.6$, and $28.7$.
The QPU circuits implement product-formula evolution and include
initial-state preparation; their resources are reported at $t=30.4$.
Binary and QPU entries count native two-qubit gates, whereas the
state-specific unary entry counts logical entangling gates before
routing.}
\label{tab:dw_compilation_routes}
\begingroup
\small
\renewcommand{\arraystretch}{1.15}
\begin{tabular*}{\linewidth}{
    @{\extracolsep{\fill}}llrr@{}
}
\toprule
Construction & Structure / processor
& 2Q count & 2Q depth \\
\midrule
Binary direct synthesis
& All-to-all structure & 95 & 94 \\
& Linear-chain structure & 156 & 152 \\
\addlinespace[4pt]
State-specific unary optimization
& Linear-chain structure & 26 & 26 \\
\addlinespace[4pt]
Unary product-formula optimization
& IBM Phoenix & 208 & 76 \\
& IonQ Forte & 182 & 50 \\
\bottomrule
\end{tabular*}
\endgroup
\end{table}

\end{appendices}
\newpage

\bibliography{main}

\end{document}